\documentclass[11pt]{article}

\usepackage{url}
\usepackage{enumerate,mathrsfs,comment}
\usepackage[algo2e]{algorithm2e}
\usepackage[]{algorithm2e}
\usepackage{enumitem}
\usepackage{amsmath}
\usepackage{natbib}
\usepackage[OT1]{fontenc}
\usepackage[hypertexnames=false, colorlinks,linkcolor=red,anchorcolor=blue,citecolor=blue]{hyperref}
\usepackage{fullpage}

\usepackage[protrusion=false,expansion=true]{microtype}
\usepackage{graphicx}

\usepackage{mathtools}
\DeclarePairedDelimiter\ceil{\lceil}{\rceil}
\DeclarePairedDelimiter\floor{\lfloor}{\rfloor}

\RequirePackage{amsmath}
\RequirePackage{amssymb}
\RequirePackage{amsthm}
\RequirePackage{bm}
\RequirePackage{url}
\usepackage{natbib}
\usepackage{multirow}
\usepackage{graphicx}
\usepackage{subfigure}
\usepackage{makecell}
\usepackage{booktabs}
\usepackage{array}
\usepackage{url}
\usepackage{algorithm}
\usepackage{algorithmic}
\usepackage{dsfont}

\newcommand{\EE}{\mathbb{E}}
\newcommand{\cD}{\mathcal{D}}
\newcommand{\diag}{{\rm diag}}

\newcommand{\pr}{{\rm{pr}}}

\newcommand{\Xtr}{X^{tr}}
\newcommand{\Ytr}{Y^{tr}}
\newcommand{\Xte}{X^{te}}

\DeclareMathOperator{\var}{{\rm var}}

\DeclareMathOperator{\bE}{{\rm E}}

\newcommand{\argmin}{\mathop{\mathrm{argmin}}}
\newcommand{\argmax}{\mathop{\mathrm{argmax}}}

\theoremstyle{plain}
\numberwithin{equation}{section}
\newtheorem{Theorem}{Theorem}[section]
\newtheorem{Lemma}[Theorem]{Lemma}
\newtheorem{Corollary}[Theorem]{Corollary}

\newtheorem{Assumption}{Assumption}

\newtheorem{lemma}{\indent \bf Lemma}

\newtheoremstyle{mytheoremstyle} %
    {\topsep}                    %
    {\topsep}                    %
    {\normalfont}                   %
    {}                           %
    {\bfseries}                   %
    {.}                          %
    {.5em}                       %
    {}  %

\theoremstyle{mytheoremstyle}

\ifx\BlackBox\undefined
\newcommand{\BlackBox}{\rule{1.5ex}{1.5ex}}  %
\fi

\ifx\QED\undefined
\def\QED{~\rule[-1pt]{5pt}{5pt}\par\medskip}
\fi

\ifx\proof\undefined
\newenvironment{proof}{\par\noindent{\bf Proof\ }}{\hfill\BlackBox\\[2mm]}
\fi

\ifx\topheorem\undefined
\newtheorem{theorem}{Theorem}
\fi
\ifx\example\undefined
\newtheorem{example}[theorem]{Example}
\fi
\ifx\property\undefined
\newtheorem{property}{Property}
\fi
\ifx\lemma\undefined
\newtheorem{lemma}[theorem]{Lemma}
\fi
\ifx\proposition\undefined
\newtheorem{proposition}[theorem]{Proposition}
\fi
\ifx\remark\undefined
\newtheorem{remark}[theorem]{Remark}
\fi
\ifx\corollary\undefined
\newtheorem{corollary}[theorem]{Corollary}
\fi
\ifx\definition\undefined
\newtheorem{definition}[theorem]{Definition}
\fi
\ifx\conjecture\undefined
\newtheorem{conjecture}[theorem]{Conjecture}
\fi
\ifx\fact\undefined
\newtheorem{fact}[theorem]{Fact}
\fi
\ifx\claim\undefined
\newtheorem{claim}[theorem]{Claim}
\fi
\ifx\assumption\undefined

\fi
\numberwithin{equation}{section}
\numberwithin{theorem}{section}

\begin{document}

\title{Domain Adaptive Bootstrap Aggregating} 

  \author
  {
  Meimei Liu\thanks{Postdoc, Department of Statistical Science, Duke University, Durham, NC, 27705. E-mail: meimei.liu@duke.edu.}    $\quad\quad$
   David B. Dunson\thanks{Professor, Department of Statistical Science, Duke University E-mail: dunson@duke.edu. Research Sponsored by the United States Office of Naval Research grant N00014-17-1-2844, and the United States National Institutes of Health grants R01-ES028804, R01ES027498-01A1.}
  }

\date{}
\maketitle

\begin{abstract}
When there is a distributional shift between data used to train a predictive algorithm and current data, performance can suffer. This is known as the domain adaptation problem. Bootstrap aggregating, or bagging, is a popular method for improving stability of predictive algorithms, while reducing variance and protecting against over-fitting. This article proposes a domain adaptive bagging method coupled with a new iterative nearest neighbor sampler.  The key idea is to draw bootstrap samples from the training data in such a manner that their distribution equals that of new testing data. The proposed approach provides a general ensemble framework that can be applied to arbitrary classifiers. We further modify the method to allow anomalous samples in the test data corresponding to outliers in the training data. Theoretical support is provided, and the approach is compared to alternatives in simulations and real data applications.
\end{abstract}

\section{Introduction}
While there is growing excitement about the accuracy of modern predictive algorithms in many domains, this excitement has been tempered by the {\em domain adaptation} problem.  In particular, certain predictive algorithms are highly sensitive to differences between the training data used to fit the algorithm and current data needing to be classified.  One may observe excellent out-of-sample predictive accuracy based on randomly splitting an initial data set, but then this accuracy can plummet when applying the classifier to new data collected under similar conditions but with a somewhat different distribution.   For example, in automatic medical diagnosis, a classifier is initially trained using data from a particular medical center or range of dates.  The classifier is meant to be used for future patients whose data may differ in subtle ways from the training patients.   Many classifiers, such as nearest neighbors, random forests, and deep neural networks, can be very sensitive to such differences, leading to poor accuracy on the new patients.                  

There is an increasing literature on addressing such domain shift problems, typically under one of two scenarios: 
(1) {\em Covariate shift} - a type of selection bias in which the marginal distribution of the covariates $X$ changes while the conditional response distribution $Y\mid X$ remains the same \citep{heckman1990varieties, cochran1973controlling, tucker2010selection}; 
 (2) {\em Prior/label/target shift} - the marginal distribution of $Y$ differs but the conditional $X\mid Y$ does not \citep{zhang2013domain, guan2019prediction, storkey2009training, lipton2018detecting}. 
 A variety of approaches have been proposed including  likelihood-based methods \citep{heckman1990varieties, chan2005word}, Bayesian meta analysis \citep{storkey2009training}, and kernel embeddings \citep{zhang2013domain}.  A common challenge of these approaches is reliance on density estimation, which is infeasible for high-dimensional and complex predictors.  Our goal is to bypass the need for density estimation.

Bootstrap aggregating, or bagging \citep{breiman1996bagging}, is routinely used for improving stability and accuracy of arbitrary base classifiers, ranging from random forests \citep{breiman2001random} to $k$-nearest neighbors \citep{hall2008choice}.  The goal of this article is to develop a domain adaptive version of bagging, referred to as DA-bagging.  The key idea is to draw bootstrap samples from the training data using a novel iterative nearest neighbor sampler to guarantee that these samples have the same distribution as new test data under the prior shift scenario discussed above. Domain adaptive bagging can also handle anomalies in the test data whose labels cannot be predicted accurately as there are no close neighbors in the training data.

\section{Domain Adaptive Bagging}\label{sec:DA-bagging}
\subsection{Notation and Preliminaries}
A classifier $C: \mathbb{R}^p \to \{1,\dots, L\}$ outputs labels $C(X) \in \{1,\dots,L\}$ for features $X \in \mathbb{R}^p$.
Suppose we have $n$ training data $\mathcal{D}_{tr} = \{(X^{tr}_1,Y^{tr}_1), \dots, (X^{tr}_n, Y^{tr}_n)\}$, and 
$m$ testing data $\mathcal{D}_{te} = (X^{te}_1,\dots, X^{te}_m)$, with each $X^{te}_i$ having a corresponding unobserved $Y^{te}_i$. 
Samples of $(X,Y) \in \mathbb{R}^p\times\{ 1,\dots, L\}$ are independent and identically distributed within the training and test groups.  
We suppose the marginal distribution of the training and testing data follow different mixture distributions as 
\begin{equation}\label{eq:model:mix}
f_{tr}(x) = \sum_{\ell=1}^L p_\ell f_\ell(x), \quad f_{te}(x) = \sum_{\ell=1}^L q_\ell f_\ell(x),
\end{equation}
where $f_\ell$, $\ell=1,\dots,L$, are shared densities of $x$ from class $\ell$, 
 $0<p_\ell\leq 1$ is the label proportion for training data satisfying $\sum_{\ell=1}^L p_\ell =1$, and $0\leq q_\ell \leq 1$ is the label proportion for test data with $\sum_{\ell=1}^L q_\ell =1$. In the literature, (\ref{eq:model:mix}) is referred to as prior/label shifting, in the sense that the prior probabilities of the classes are different, 
 but the conditional feature distributions are shared. 

\subsection{Methodology}\label{sec:DA-bagging}
Key to our proposed domain adaptive bagging algorithm is a novel iterative nearest neighbor sampler to generate bootstrap samples from training data
$\mathcal{D}_{tr}$ with the guidance of test data $\mathcal{D}_{te}$, so that the samples are equal in distribution to the test data.
Let $\|\cdot \|$ be a metric defined on the separable metric space $\mathbb{R}^p$. 
For any $X^{te} \sim f_{te}(x)$, we reorder the samples in $\mathcal{D}_{tr}$ as 
$
\big\{X_{(1)}(X^{te}), Y_{(1)}(X^{te})\big\}, \dots, \big\{X_{(n)}(X^{te}), Y_{(n)}(X^{te})\big\}, 
$ such that 
$$
\|X^{te}- X_{(1)}(X^{te})\| \leq \dots \leq \|X^{te}- X_{(n)}(X^{te})\|,
$$
and define its $k$-nearest neighbors in $\mathcal{D}_{tr}$ as $\mathcal{N}_{X^{te}}^{k}=\big\{X_{(1)}(X^{te}), \dots, X_{(k)}(X^{te})\big\}$.  

For each $X_j^{te}$ ($j=1,\dots,m$), we obtain a stratified bootstrap sample $\mathcal{D}_{tr,j}^{(1)}$ of size  $\ceil{n/m}$  from $\mathcal{N}_{X_j^{te}}^k$ using 
the $L$ classes as strata with weights $\pi_j = (\pi_{j1},\ldots,\pi_{jL})$ where 
\begin{equation}\label{eq:bootstrap}
\pi_{j, \ell} = \frac{1}{k}\#\big\{\Ytr_i = \ell\; |\; \Xtr_i \in \mathcal{N}_{\Xte_j}^{k} , i=1,\dots,n.\big\} \quad \mbox{for } \ell=1,\dots,L.
\end{equation} 
Repeating for $j=1,\dots, m$, we have new training data $\mathcal{D}_{tr}^{(1)}=(\cD_{tr,1}^{(1)},\dots, \cD_{tr,m}^{(1)})$.
We repeat the above procedure to obtain $\mathcal{D}_{tr}^{(2)}$ and so on for $T$ iterations to obtain $\mathcal{D}_{tr}^{(T)}$.   We stop iterating when the proportions of observations within each class differ from $\mathcal{D}_{tr}^{(T-1)}$ to  $\mathcal{D}_{tr}^{(T)}$ by less than a small threshold.  

Figure \ref{fig:example} (a) provides an illustrative example. In Figure \ref{fig:example} (a), blue and green dots represent data points in class $\{1\}$ and class $\{2\}$. For each point in the test data, shown as black circles in the training data, we find its nearest neighbor in the training data and draw $\ceil{n/m}$ samples from class $\{\ell\}$, where $\ell$ is the label of its nearest neighbor. In the sampled data plot, larger circles correspond to repeated data points.  Figure \ref{fig:example} (b) shows histograms of the training data, test data and first to fifth samples.
\begin{figure}\label{fig:example}
\center
\includegraphics[width=0.90\textwidth]{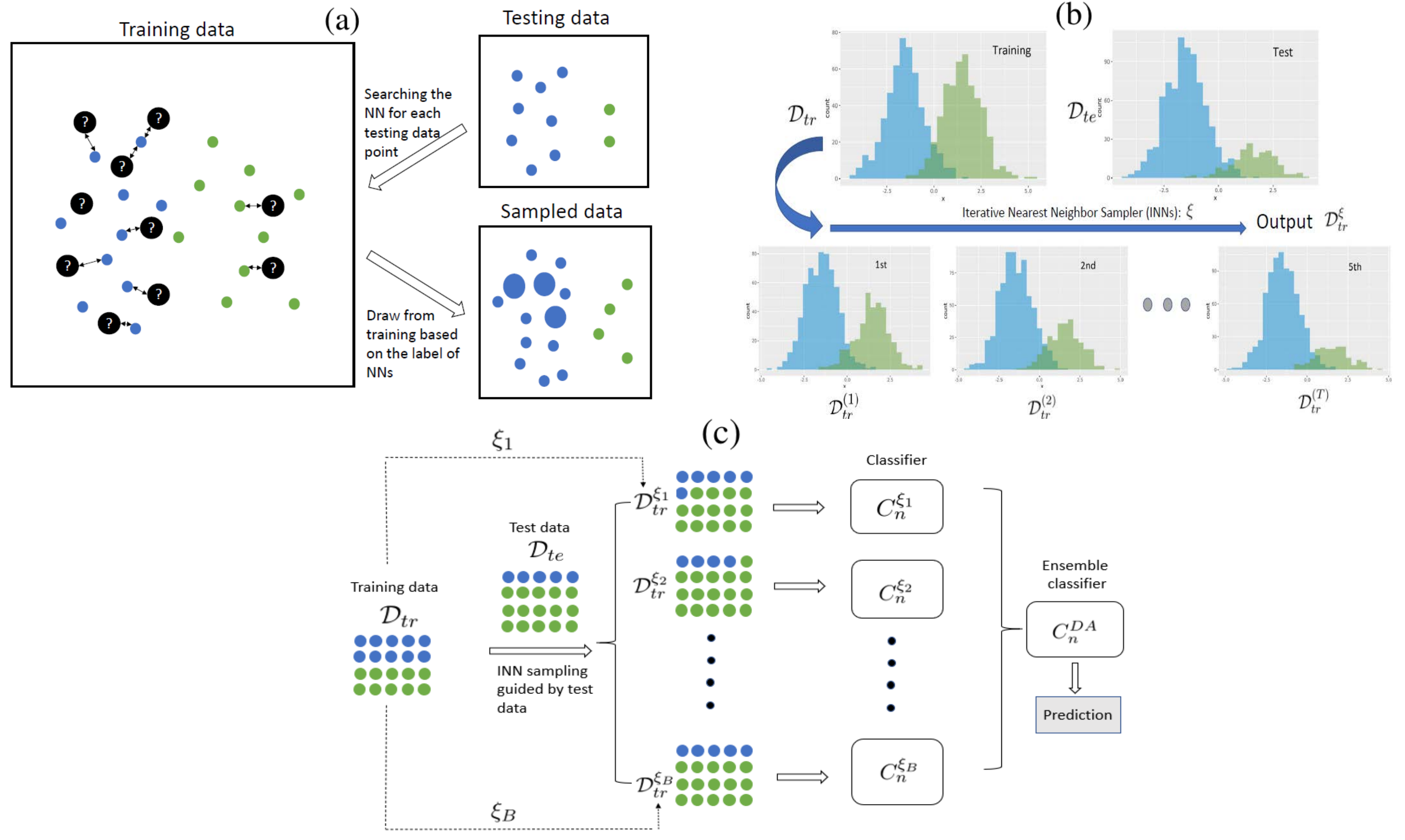}\\ 
\caption{(a) Flowchart of the nearest neighbor sampler. (b) Illustrative figure of the histogram of samples by iterative nearest neighbor sampler. (c) Flowchart of the domain adaptive sampling schedule. }\label{fig:example}
\end{figure}

Let $\xi$ be a randomization parameter controlling the mapping from the initial $\mathcal{D}_{tr}$ to the sampled $\mathcal{D}_{tr}^{(T)}$ under the above procedure. 
 Implementing the procedure $B$ times, we obtain randomizations $\xi_1,\dots, \xi_B$ and corresponding datasets $\cD_{tr}^{\xi_1}, \dots, \cD_{tr}^{\xi_B}$. 
Parameters $\xi_1,\dots, \xi_B$ are conditionally independent and identically distributed given $\mathcal{D}_{tr}$ and $\mathcal{D}_{te}$. 
Based on each $\cD_{tr}^{\xi_b}$, we construct a base classifier 
  $
  C^{\xi_b}_n(X)
   = C_n(X; \mathcal{D}_{tr}, \mathcal{D}_{te}, \xi_b).
  $ 
Our domain adaptive bagging ensemble classifier $C_n^{DA}$ is 
\begin{equation}\label{eq:voting}
C_n^{DA}(X) =\argmax_\ell \#\{b\in \{1,\cdots, B\} :   C^{\xi_b}_n (X)=\ell\}.
\end{equation}
Figure \ref{fig:example} (c) shows the flowchart of the domain adaptive sampling schedule. The detailed steps are summarized in Algorithm 1.

 \begin{algorithm}[h]\label{alg:rbagging}
1. \For{Parallel $b=1, \dots, B$}{ 
		\For{Parallel $t=1,\dots, T-1$}{
   			 \For{Parallel $j=1,\dots, m$}{
(a) For $X_j^{te}$, find its $k$-nearest neighbor in $\mathcal{D}_{tr}^{b*,(t)}$, denoted as $\{\mathcal{N}_{\Xte_i}^{k}\}$. \\
(b) Define $\pi_{j} = (\pi_{j,1},\dots, \pi_{j,L})$ with $\pi_{j,\ell}$ defined as 
$$
\pi_{j, \ell} = \frac{1}{k}\#\big\{\Ytr_i = \ell\; |\; \Xtr_i \in \mathcal{N}_{\Xte_j}^{k} \big\} \quad \mbox{for } \ell= 1,\dots,L.
$$
(c) Sample $(n_{j1},\dots, n_{jL})$ from Multinomial$(\ceil{n/m}, \pi_{j})$.\\
(d) For $\ell= 1,\cdots, L$, draw $n_{j\ell}$ bootstrap samples with replacement from the $\ell$-th class in $\mathcal{D}_{tr}^{b*,(t)}$. 
}
Denote the $\{n_{j\ell}\}$ samples together as $\mathcal{D}_{tr}^{b*,(t+1)}$. 
}
Define $\mathcal{D}_{tr}^{b*,(T)}$ as $\mathcal{D}_{tr}^{\xi_b}$, and construct the classifier $C^{\xi_b}_n(\cdot) = C_{n,\cD_{tr}^{\xi_b}}(\cdot)$.
}

2. Build the ensemble classifier as the majority voting in (\ref{eq:voting}).
  \caption{Domain adaptive bagging }
\end{algorithm}

\section{Theoretical Results}\label{sec:theory}
\subsection{Consistency of resampling algorithm}
We study properties of domain adaptive bagging using $k=1$ in Algorithm 1. To simplify the presentation, we consider binary classification with labels equal to $1$ or $2$. Then model (\ref{eq:model:mix}) can be written as  
\begin{equation}\label{eq:model:two_class}
f_{tr}(x) = p_1 f_1(x) + p_2f_2(x)  \quad \quad\textrm{and} \quad \quad f_{te}(x) = q_1f_1(x) + q_2 f_2 (x), 
\end{equation}
where $p_1 = \pr(Y^{tr} =1)$, $p_2= 1-p_1 =\pr(Y^{tr} =2)$,  $q_1= \pr(Y^{te} = 1)$ and $q_2%
= \pr(Y^{te}=2)$.

We aim to show that the bootstrapped data $\cD_{tr}^{\xi_1}, \dots, \cD_{tr}^{\xi_B}$ can represent the testing data with probability 1. Further, we establish the generalization error of general classifiers based on domain adaptive bagging, and characterize the algorithmic convergence. 

Letting $k=1$ and $L=2$ in Algorithm 1, the steps can be simplified as in Algorithm 2. 
For each $X_j^{te}\in \mathcal{D}_{te}$, if its nearest neighbor in $\mathcal{D}_{tr}$ is from class $\ell \in \{1,2\}$, we randomly sample $\ceil{n/m}$ data from class $\ell$ with replacement. Repeat this procedure for all $m$ testing samples to obtain data $\mathcal{D}_{tr}^{1*,(1)}$, where $1*$ represents the $1$st bootstrap copy, and $(1)$ means the first iteration in that bootstrap copy.   Viewing $\mathcal{D}_{tr}^{1*,(1)}$ as the new training data, we repeat this procedure, obtaining $\mathcal{D}_{tr}^{1*,(t)}$ at the $t$th iteration.  In Theorem \ref{thm:sample:dist}, we show that for $X\in \mathcal{D}_{tr}^{1*,(t)}$, the corresponding density $f^{(t)}(x)$ approaches $f_{te}(x)$ as $t$ increases. 

 \begin{algorithm}[h]\label{alg:rbagging:1nn}

1. \For{Parallel $b=1, \dots, B$}{ 
      \For{Parallel $t=1,\dots, T-1$}{
    \For{Parallel $j=1,\dots, m$}{
(a) For $X_j^{te}\in \mathcal{D}_{te}$, find its nearest neighbor in $\mathcal{D}_{tr}^{b*,(t)}$, denoted as $X_{(1)}$, and the corresponding label as $Y_{(1)}$. \\
(b) Bootstrap $n_{j}= \ceil{n/m}$ samples with replacement from the $\ell$-th class in $\mathcal{D}_{tr}^{b*,(t)}$, where $\ell = Y_{(1)}$. 
}
Collect the $n_{1},\dots, n_{m}$ samples together as $\mathcal{D}_{tr}^{b*,(t+1)}$. 
}
Denote $\mathcal{D}_{tr}^{b*,(T)}$ as $\mathcal{D}_{tr}^{\xi_b}$. Construct the classifier $C^{\xi_b}_n(\cdot) = C_{n,\mathcal{D}_{tr}^{\xi_b}}(\cdot)$.
}
2. Ensemble the classifiers based on $\mathcal{D}_{tr}^{\xi_1},\cdots, \mathcal{D}_{tr}^{\xi_B}$ as the majority voting 
$$C_n^{DA}(X) =\argmax_\ell \#\{b\in \{1,\cdots, B\} :   C^{\xi_b}_n(X)=\ell\}.$$ 
\caption{DA-bagging based on 1-NN with $L=2$} 
\end{algorithm}

Let $X^{te} \sim f_{te}(x)$ with $\{X_i\}_{i=1,\dots,n}$ independent and identically distributed from $f_{tr}(x)$.  
Denote the nearest neighbor of $X^{te}$ in $\{X_1,\cdots, X_n\}$ as $X_{(1)}(X^{te})$ so that
$
\min_{i\in\{ 1,\dots, n\}} \|X_i-X^{te}\| = \|X_{(1)}(X^{te})- X^{te}\|.
$                   
We denote the corresponding label of $X_{(1)}(X^{te})$ as $Y_{(1)}(X^{te})$. 
We first show that, under the distributional shift from $f_{tr}$ to $f_{te}$ defined in (\ref{eq:model:mix}), for any $X^{te}\in \mathcal{D}_{te}$, its nearest neighbor $X_{(1)}(X^{te})$ in $\mathcal{D}_{tr}$ converges to $X^{te}$ with probability one. 
\begin{Lemma}\label{le:convergence}
Suppose $f_{tr}(x)$ and $f_{te}(x)$ follow $(\ref{eq:model:mix})$. Let $X^{te} \sim f_{te}(x)$ with $\{X_i\}$ independent and identically distributed from $f_{tr}(x)$. 
 Then $X_{(1)}(X^{te})\to X^{te}$ with probability 1.  
\end{Lemma}

\begin{Theorem}\label{thm:sample:dist}
Consider model $(\ref{eq:model:two_class})$. Assume $\pr_{f_1} \big( X: f_1(X)= f_2(X)\big) = 0$. Under Algorithm S1, denote $\mathcal{D}_{tr}^{\xi_b} = \big\{ (X^{\xi_b}_1, Y^{\xi_b}_1), \dots, (X^{\xi_b}_n, Y^{\xi_b}_n)\big\}$. Denote the 
conditional density of $X_i^{\xi_b} \mid Y_i^{\xi_b}=\ell$ as $f^{(T)}_\ell(x)$, and the marginal distribution of $X^{\xi_b}_i$ as $f^{(T)}$. Then 
 $\{(X_i^{\xi_b}, Y_i^{\xi_b})\}$ are independent and identically distributed,  
$\pr(Y_i^{\xi_b}=\ell)= q_\ell$, $f^{(T)}_\ell(x)=f_\ell(x)$, and hence, 
$
f^{(T)}(x) = \lim_{t\to \infty} f^{(t)} (x) = \sum_{\ell=1}^L q_\ell f_\ell(x). 
$
\end{Theorem}
Here ${\pr}_{f_1}$ refers the probability measure of $X\sim f_1(x)$. 
Theorem \ref{thm:sample:dist} shows that each sample in the bootstrap dataset $\cD_{tr}^{\xi_b}$ follows the same distribution as the testing data, that ${\pr}(Y^{te}= \ell)= q_\ell$, $X^{te}\mid Y^{te}= \ell \sim f_\ell$ and the 
 marginal density $f_{te}(x)= \sum_{\ell=1}^L q_\ell f_\ell(x)$. 

\subsection{Prediction error based on domain adaptive bagging} 
 In this section, we analyze the testing error of basic classifiers based on domain adaptive bagging. To distinguish between different sources of randomness, we denote $\pr(\cdot\mid \mathcal{D}_{tr}, \mathcal{D}_{te}), \bE(\cdot\mid \mathcal{D}_{tr}, \mathcal{D}_{te})$ as the probability and expectation respectively, taken over the randomness from $\xi_1,\dots, \xi_B$ conditional on the observed training and test data $(\mathcal{D}_{tr}, \mathcal{D}_{te})$. We define $\pr, \bE$ as the probability and expectation taken over all random quantities.

Considering model (\ref{eq:model:two_class}), the test error of a classifier $C$ is defined as 
\begin{equation}\label{eq:test:error}
R(C) : = q_1 \int \mathds{1}_{\{C(x)= 2\}} f_1(x)dx + q_2 \int \mathds{1}_{\{C(x)=1\}} f_2(x) dx.
\end{equation}
$R(C)$ is minimized by the Bayes classifier defined as 
$C^{Bayes} (x) : = 1$ if $\eta(x)\geq 1/2$ and $2$ otherwise, 
where $\eta(x) = {\pr} (Y^{te}=1 | X^{te} = x) = q_1 f_1(x)/\{q_1f_1(x) + q_2 f_2(x)\}$. The corresponding Bayes risk for $C^{Bayes}$ is 
 $$R(C^{Bayes}) = \EE_{X\sim f_{te}} [\min\{\eta(X), 1-\eta(X)\}] = \int \min\{q_1 f_1(x), q_2 f_2(x)\} dx.$$

Given a base classifier $C$ and the new training samples $\cD_{tr}^{\xi_b}$ ($b=1,\dots, B$) obtained in Algorithm 2, we have a sequence of trained classifiers $C_n^{\xi_1},\dots, C_n^{\xi_B}$. We define $\Lambda_n(x) = \frac{1}{B}\sum_{b=1}^B \mathds{1}_{\{C_n^{\xi_b}(x)=1\}}$, and the ensemble classifier $C_n^{DA}$ based on domain adaptive bagging as 
$C_n^{DA} (x) = 1$ if $\Lambda_n(x)\geq 1/2$ and $0$ otherwise. 
In Theorem \ref{thm:pred:error}, we show the test excess risk, the difference between expected test error of $C_n^{DA}$ and the Bayes risk, can be controlled by the expected test excess risk of the classifier  $C_n^{\xi_b}$ based on a single sample $\xi_b$.  

\begin{Theorem}\label{thm:pred:error}
Assume model (\ref{eq:model:two_class}) holds. Based on Algorithm S1, 
for $1\leq b \leq B$, we have 
\begin{equation}\label{eq:pred:bound}
\bE \{R(C_n^{DA})\} - R(C^{Bayes}) \leq 2 [\bE \{R(C_n^{\xi_b})\} - R(C^{Bayes})]. 
\end{equation}
\end{Theorem}
 
Theorem \ref{thm:pred:error} provides a general bound for different base classifiers. 
By Theorem \ref{thm:sample:dist}, the marginal distribution of $X_i^{\xi_b}$ follows $f_{te}(x)$ defined in (\ref{eq:model:two_class}). By definition (\ref{eq:test:error}), calculating $R(C^{Bayes})$ only involves the density function $f_{te}$. Therefore, viewing $\cD_{tr}^{\xi_b}$ as the training data, the distributional shift is removed from the 
 upper bound in (\ref{eq:pred:bound}).  Hence, existing results on performance of base classifiers can 
 be used to provide explicit bounds on the expected test excess risk. 

 In Corollary \ref{cor:knn}, we bound the excess risk in (\ref{eq:pred:bound}) using $k$-nearest neighbors as the base classifier. 
  Given $X^{te}\in\mathbb{R}^p$ generated from the density $f_{te}(x)$, we first order the data in $\cD_{tr}^{\xi_b}$ as $(X^{\xi_b}_{(1)}, Y^{\xi_b}_{(1)}), \dots, (X^{\xi_b}_{(n)}, Y^{\xi_b}_{(n)})$ such that 
  $
  \|X^{\xi_b}_{(1)}-X^{te}\|\leq \dots \leq \|X^{\xi_b}_{(n)} - X^{te}\|$, with ties split at random. 
The $k$-nearest neighbor classifier is defined as 
$C_n^{\xi_b}(X^{te}) = 1$ if $\frac{1}{k}\sum_{i=1}^k \mathds{1}_{\{Y^{\xi_b}_{(i)}=1\}}\geq 1/2$ and $2$ otherwise. 
\cite{hall2008choice} established the rate of convergence of the excess risk with the optimal choice of $k$. Combining with Theorem \ref{thm:pred:error}, we have the following Corollary.

\begin{Corollary}\label{cor:knn}
Suppose $X\in \mathbb{R}^p$ is a random variable with density $f_{te}(x)$ in $(\ref{eq:model:two_class})$. Under regularity conditions, if $k$ is chosen as $O(n^{4/(p+4)})$,  we have 
$$
\bE \{R(C_n^{DA})\} - R(C^{Bayes})  \leq 2 \big[\bE \{R(C_n^{\xi_b})\} - R(C^{Bayes})\big] = O(n^{-4/(p+4)}). 
$$
\end{Corollary}

We next focus on the algorithmic randomness introduced by $\xi_b$ given the observations $(\mathcal{D}_{tr}, \mathcal{D}_{te})$. 
Algorithmic convergence has been studied for randomized ensembles to analyze the effect of the ensemble size $B$ on prediction error; see \cite{cannings2017random} and \cite{lopes2020estimating}. Define $\mu_n (X^{te})  = \bE\{\Lambda_n(X^{te})\mid \mathcal{D}_{tr}, \mathcal{D}_{te}\} = \pr\{C_n^{\xi_1}(X^{te})=1 \mid \mathcal{D}_{tr}, \mathcal{D}_{te}\}$. Intuitively, $\mu_n (X^{te})$ represents infinite bootstrap samples with $B\to \infty$. Define the classifier with infinite ensemble size as  
$C_n^{DA*}(X^{te}) = 1$ if $\mu_n(X^{te})\geq 1/2$ and $2$ otherwise. 
In Theorem \ref{thm:alg}, we characterize 
how the test error of $C_n^{DA}$ based on an ensemble of size $B$ converges to the ideal level of an infinite ensemble of $C_n^{DA*}$, in terms of the algorithmic randomness introduced by $\xi_b$. 

We first introduce an assumption regarding the distribution of $\mu_n(X^{te})$. 
 Define the distribution functions of $\mu_n(X^{te})$ conditional on $Y^{te}$ as  $\mathcal{L}_{\mu_n,\ell}( t \mid \mathcal{D}_{tr}, \mathcal{D}_{te}, Y^{te}=\ell)$ for $\ell =1,2$. That is, 
 \begin{align*}
 \mathcal{L}_{\mu_n,1}( t \mid \mathcal{D}_{tr}, \mathcal{D}_{te}, Y^{te}=1) = &\pr[\{X^{te}\in \mathbb{R}^p: \mu_n(X^{te})\leq t\} \mid \mathcal{D}_{tr}, \mathcal{D}_{te}, Y^{te}=1], \\
  \mathcal{L}_{\mu_n,2}( t \mid \mathcal{D}_{tr}, \mathcal{D}_{te}, Y^{te}=2) = &\pr[\{X^{te}\in \mathbb{R}^p: \mu_n(X^{te})\leq t\} \mid \mathcal{D}_{tr}, \mathcal{D}_{te}, Y^{te}=2]. 
 \end{align*}

\begin{Assumption}\label{assump:random}
For $\ell\in\{1,2\}$, $\mathcal{L}_{\mu_n,\ell}( t \;|\; \mathcal{D}_{tr}, \mathcal{D}_{te}, Y^{te}=\ell)$ 
is twice differentiable at $t= 1/2$. 
\end{Assumption}

\begin{Theorem}\label{thm:alg}
Under model (\ref{eq:model:two_class}), Algorithm S1, and Assumption \ref{assump:random}, as $B\to \infty$,  
 \begin{align*}
 \bE\{R(C_n^{DA})\mid \mathcal{D}_{tr}, \mathcal{D}_{te}\} - R(C_n^{DA*}) = & \frac{\gamma_n}{B} + o(B)\\
 \lim_{B\to \infty} B \var\{R(C_n^{DA})\mid \mathcal{D}_{tr}, \mathcal{D}_{te}\} \leq 1/4\bar{g}_n^2(1/2),
 \end{align*}
 where 
$\gamma_n = \{1/2 - (B/2 - \floor{B/2})\}\{q_1 g_{n,1}(1/2) - q_2 g_{n,2}(1/2)\}+ \frac{1}{8} \{q_1 \dot{g}_{n,1}(1/2)-q_2 \dot{g}_{n,2}(1/2)\}$, and $\bar{g}_n(1/2) = q_1 g_{n,1}(1/2) + q_2 g_{n,2}(1/2)$. Here $g_{n,\ell}$ and $ 
\dot{g}_{n,\ell}$ are the first and second order derivative of $\mathcal{L}_{\mu_n,\ell}( t \mid \mathcal{D}_{tr}, \mathcal{D}_{te}, Y^{te}=\ell)$ for $\ell=1,2$.  
\end{Theorem} 
Theorem \ref{thm:alg} shows that the bias and variance of the test error are of order $O(1/B)$.   
The proof of Theorem \ref{thm:alg} follows from Theorem 1 in \cite{cannings2017random}.

 \section{Domain adaptive bagging with anomalies }\label{sec:robust_DA-bagging}
In this section, we consider the situation in which anomalous samples are present in the testing data, with `anomalous' meaning that these samples would be considered as outliers if they were observed in the training sample.  In particular, we consider the model 
 \begin{equation}\label{eq:model:outlier}
 f_{tr}(x) = \sum_{\ell=1}^L p_\ell f_\ell(x), \quad f_{te}(x) = \sum_{\ell=1}^L q_\ell f_\ell(x) + \epsilon f_{out}(x),
 \end{equation}
where $p_\ell>0$ satisfying $\sum_{\ell=1}^L p_\ell =1$ and $q_\ell \geq 0, \epsilon\geq 0$ satisfying $\sum_{\ell=1}^L q_\ell + \epsilon =1$. Denote $\pr_\ell$ as the probability measure of $X\sim f_\ell(x)$ for $\ell=1,\dots, L$, and $\pr_{0}$ as the probability measure of $X\sim f_{out}(x)$. 
 This model allows not only changes in the mixture proportions between training and test but also an additional mixture component for the test data corresponding to anomalous observations that may be dissimilar to any of the training samples.
 
We first detect the anomalies before conducting domain adaptive resampling.  
Denote $\mathcal{D}_{tr}= (\mathcal{D}_{tr,1},\dots, \mathcal{D}_{tr,L})$, where $\mathcal{D}_{tr,\ell}$ only contain the training data with label $\{\ell\}$, having sample size $|\mathcal{D}_{tr,\ell}|= n_\ell$. For any $x$, define the squared distance between $x$ and its $k$-nearest neighbors in $\mathcal{D}_{tr,\ell}$ as 
 \begin{equation}\label{eq:emp:dtm}
\widehat{d}^2_\ell(x) = \frac{1}{k}\sum_{X_i \in \mathcal{N}_{x}^{k,\ell}} \|X_i - x\|^2,
\end{equation}
where $\mathcal{N}_{x}^{k,\ell}$ are the $k$-nearest neighbors of $x$ in $\mathcal{D}_{tr,\ell}$. 
Intuitively,  
if $x$ is an anomaly, then for any $\ell\in \{1,\dots, L\}$, $\widehat{d}^2_\ell(x)$ is large. 
A population version of $\widehat{d}^2_\ell(x)$ is called the distance to measure \citep{chazal2011geometric,chazal2017robust} defined as 
\begin{equation}\label{eq:dtm}
d^2_\ell(x) = \frac{1}{m_\ell} \int_0^{m_\ell} r_{\ell,t}(x) dt,
\end{equation}
where $m_\ell=k/n_\ell$ is called resolution and $r_{\ell,t}(x) = \argmin\{r: \pr_\ell\big(\|X-x\|^2 \leq r\big)>t\}$.  By equation (\ref{eq:dtm}), a relatively large distance to measure happens in two situations: (1) $x$ is a tail sample from $f_\ell(x)$; (2) $x$ is an anomaly. 
Letting $\pr_{n_\ell}$ be the empirical probability measure that puts mass $1/n_{\ell}$ on each $X_i\in\mathcal{D}_{tr,\ell}$,  the distance to measure $\pr_{n_\ell}$ at resolution $m_\ell$ is exactly equation (\ref{eq:emp:dtm}). 

Observing this, for each $X^{te}\sim f_{te}(x)$, we construct a statistic for anomaly detection as 
\begin{equation}\label{eq:test_anomaly}
T(X^{te}) = \prod_{\ell=1}^L \mathds{1}_{\{\widehat{d}_\ell (X^{te}) > c_\ell\}},
\end{equation}
where $c_\ell$ is a constant threshold that will be specified later. 
In Lemma S3, 
we show that $\widehat{d}_\ell (x)$ is a consistent estimator of $d_\ell(x)$. Then for any $X^{te}\in \mathcal{D}_{te}$, $T(X^{te})=1$ if and only if $d_\ell(X^{te})>c_\ell$, that is,  $\widehat{d}_\ell(X^{te})>c_\ell$, for all $\ell\in \{1,\dots, L\}$. 

Given $X^{te}\in \mathcal{D}_{te}$ with unobserved label $Y^{te}$, we propose a detection rule as $\Psi(X^{te})=\mathds{1}_{\{T(X^{te})>c\}}$ for $0<c<1$. 
The type I error is the probability of wrongly detecting the anomaly, $\pr\{T(X^{te})>c \mid Y^{te}\in \{1,\dots, L\}\}$. 
The power of $\Psi(X^{te})$ is  $\pr\{T(X^{te})>c \mid Y^{te}\not\in \{1,\dots, L\}\}.$ 
In Theorem \ref{thm:test}, we show that the type I error can be controlled at a nominal level $\alpha$ by properly choosing thresholds $c_\ell$s, while guaranteeing high power. Before the formal statement of Theorem \ref{thm:test}, we first state some assumptions regarding the distribution of the distance to measure and separation between the normal and abnormal samples.

\begin{Assumption}\label{assump:1}
\begin{enumerate}[label=(\alph*)]
\item For each $\ell= 1,\dots, L$, given a nominal level $\alpha$ where $0< \alpha <1$, there exists a positive finite constant $c_\ell$ satisfying $c_\ell= \argmin[ c: \pr_\ell\{X: d_\ell(X) \leq c\}\geq 1-\alpha ]$. 
\label{assump:a}
\item 
For any $\delta\in (0,1)$, denote $\beta_{n\ell} = \sqrt{(4/n_\ell)\{(p+1)\log(2n_\ell) + \log (8/\delta)\}}$. There exists a constant $M$ satisfying $M \geq 2 c_\ell + C \beta_{n\ell} (\beta_{n\ell} + \sqrt{m_\ell})$, such that given the anomaly $x_{out}$ sampled from $\pr_0$, $\pr_\ell(\|X-x_{out}\|^2 \leq M) \leq \epsilon$, where $C$ is a constant and  $\epsilon < m_\ell/2$. \label{assump:b}
\end{enumerate}
\end{Assumption}  

Assumption \ref{assump:1} \ref{assump:a} defines a safety zone for $X\sim \pr_\ell$, that is,  $\mathcal{A}_\ell = \{X: X\sim f_\ell(x), \; d_\ell(X) \leq c_\ell\}$. Based on $\mathcal{A}_\ell$, Assumption \ref{assump:1} \ref{assump:a} separates $X\sim \pr_\ell$ into two parts based on the distance to measure being smaller or larger than 
 $c_\ell$. 
  Assumption \ref{assump:1} \ref{assump:b} requires that the distance between anomalies sampled from $f_{out}$ and samples from $f_\ell$ can be lower bounded with large probability.

\begin{Assumption}\label{assump:2} 
(a) There exist positive constants $C = C (\pr_\ell)$ and $\varepsilon_0 = \varepsilon_0(\pr_\ell)$, such that for all $0< \varepsilon < \varepsilon_0$ and $\eta \in \mathbb{R}$, for any $x$, and $0< t < m$, 
$
\pr_\ell \big\{\|X-x\|\leq r_{\ell,t}(x) + \eta \big\} - \pr_\ell \big\{ \|X-x\| \leq r_{\ell,t}(x) \big\}\leq \varepsilon,
$
 $|\eta| < C \varepsilon $. 
(b) For any $x \sim \pr_\ell$, if $\kappa < C \varepsilon$, then 
$
\pr_\ell \big\{ d_\ell (x) > c_\ell - \kappa\} - \pr_\ell\{d_\ell(x) > c_\ell\big\} \leq \varepsilon. 
$
\end{Assumption}
Assumption \ref{assump:2} implies that $\pr_\ell$ has non-zero probability around the boundary of 
the ball centered at $x$ with radius $r_{\ell,t}(x)$.

\begin{Theorem} \label{thm:test}
Consider model (\ref{eq:model:outlier}). Suppose Assumptions \ref{assump:1} and \ref{assump:2} hold. 
\begin{enumerate}[label=(\alph*)]
  \item As $n\to \infty$, for $X^{te}\in \mathcal{D}_{te}$ with unobserved label $Y^{te}\in \{1,\dots, L\}$, for $0<c<1$, $\pr\{T(X^{te})>c \mid Y^{te}\in\{1,\dots, L\}\} 
   \leq \alpha$.\label{thm:test:size} 
  \item For $X^{te}\in \mathcal{D}_{te}$, $0<c<1$, we have ${\pr}\{T(X^{te})>c \mid  \; Y^{te}\not\in \{1,\dots, L\} \} \geq 1-L\delta $, where $\delta$ is specified in Assumption \ref{assump:1} \ref{assump:b}.  
  \label{thm:test:power}
\end{enumerate}
\end{Theorem}

In practice, we cannot directly calculate $c_\ell$ since it depends on the unknown density $f_\ell$.  Instead, we use a data splitting method summarized in Algorithm \ref{alg:anomaly} to approximate the cutoff and detect anomalies. These anomalies are removed before applying domain adaptive bagging.

 \begin{algorithm}[h!]\label{alg:anomaly}
Input: labeled training data $D_{tr,\ell}$, testing data $\mathcal{D}_{te}$, $k$, $\alpha$ : \\ 
1. \For{Parallel $\ell=1, \dots, L$}{ 
    Split $D_{tr,\ell}$ as $D_{tr,\ell}^{(1)}$ and $D_{tr,\ell}^{(2)}$. \\
      \For{Parallel $i=1,\dots, |D_{tr,\ell}^{(1)}|$}{
       For each $z_i \in D_{tr,\ell}^{(1)}$, find its $k$-NN in $D_{tr,\ell}^{(2)}$ denoted as $\mathcal{N}_{z_i}^k$, and calculate 
       $$
       \widehat{d}^2_\ell(z_i) = \frac{1}{k} \sum_{X_j\in \mathcal{N}_{z_i}^k}\|X_j - z_i\|^2.
       $$ 
}
Set $\widehat{c}_\ell$ as the $(1-\alpha)$-th quantile of $(\widehat{d}_\ell(z_1),\dots, \widehat{d}_\ell(z_{|D_{tr,\ell}^{(1)}|}))$. 
}
2. \For{$i=1,\dots, |\mathcal{D}_{te}|$}{
Find the $k-$NN of $X_i^{te}\in \mathcal{D}_{tr}$ denoted as $\mathcal{N}^k_{X_i^{te}}$, and calculate 
$$
\widehat{d}_\ell^2(X_i^{te}) = \frac{1}{k}\sum_{X_j\in \mathcal{N}^k_{X_i^{te}}} \|X_j - X_i^{te}\|^2. 
$$
Calculate the test statistics $T(X_i^{te}) = \prod_{\ell=1}^L \mathds{1}_{\{\widehat{d}_\ell (X_i^{te})> \widehat{c}_\ell\}}$.\\
Set $X_i^{te}\in \mathcal{D}_0$ if $T(X_i^{te})= 1$. 
}

  \caption{Anomaly Detection}

\end{algorithm}

\section{Simulation Study}\label{sec:simulation}
\subsection{An illustrative example} 
We first consider a toy example to compare domain adaptive and classical bagging when a distribution shift occurs.
We generate training data with sample size $n=300$ from 
$f_{tr}(x,y) = \pr(Y^{tr}=y)f(x| Y=y) = \frac{1}{3} f_y(x)$ for $y=1,2,3$, with $f_1$ the density function of $N((1,1)^\top, \Sigma)$, $f_2$ the density of $N((1,4)^\top, \Sigma)$, and $f_3$ the density of $N((1,7)^\top, \Sigma)$. We set $\Sigma$ as a $2\times 2$ matrix with diagonal entries 1 and off-diagonal entries $0.2$. 
We generate testing data with sample size $m=300$ 
from $f_{te}(x,y) = \pr(Y^{te}=y)f(x| Y=y) = \pr(Y^{te}=y) f_y(x)$ with $\pr(Y^{te}=1)= \pr(Y^{te}=2)= 10/21$ and $\pr(Y^{te}=3)= 1/21$. 
Hence, the test data have different class proportions. We use multinomial regression as the base classifier. 

\begin{figure}[h!]
\centering
\includegraphics[width=1.00\textwidth]{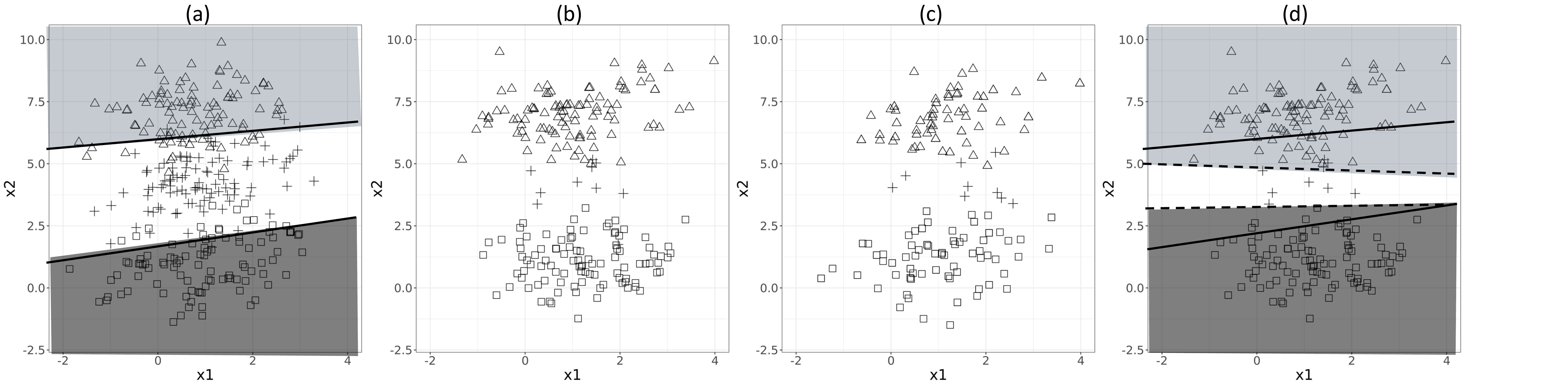}
\caption{ (a) training data; (b) testing data; (c) bootstrap samples based on INNs in Algorithm 1 with $k=5$; (d) the dashed lines are the decision boundaries based on DA-bagging; and the solid lines are the averaged decision boundaries from classic bagging; the colored region are the decision regions based on Bayes classifier. } \label{figure:exp1}
\end{figure}

We implement domain adaptive bagging using Algorithm 1 with $k=5$ and $B=10$.
Figure \ref{figure:exp1} (a) and (b) show the generated training and testing data. Figure \ref{figure:exp1} (c) shows one replicate of the iterative nearest neighbor sampler with 
$T=5$. Clearly, the sampled data approximates the testing data. 
We further compare the decision boundaries. The solid lines in Figure \ref{figure:exp1} (d) are the averaged decision boundaries via classical bagging; clearly performance is suboptimal. The colored regions are the decision regions based on the Bayes classifier assuming the distribution of $\pr(Y^{te}|X=x)$ is known. 
The dashed lines in Figure \ref{figure:exp1} (d) are learned from domain adaptive bagging, and are closer to the Bayesian rule for the testing data. 
\subsection{Simulation study without outliers}\label{sec:sim:inlier}
In the simulated experiments, we assess the empirical performance of domain adaptive bagging combined with four base classifiers including logistic regression, classification and regression trees , random forests, and linear discriminant analysis. 
We compare our method with two popular domain adaptation methods under the above four base classifiers. The first is kernel mean matching  \citep{zhang2013domain}. 
The second is balanced and conformal optimized prediction sets \citep{guan2019prediction}. 
 We also compare with the base classifiers without considering domain adaptation. 
 We use classification and regression trees as the default classifier in bagging. We keep the tuning parameters for each classifier the same when coupled with different data adaptation approaches. For example, we fix the number of trees and variables to possibly split at in each node in the random forest classifier for different methods. In domain adaptive bagging, we stop iterating when the class proportions differ from $\mathcal{D}_{tr}^{(T-1)}$ to  $\mathcal{D}_{tr}^{(T)}$ by no more than $0.01$. We set $B=500$ for both bagging methods. 

We consider two design scenarios. In each scenario, we generate the training data from two classes: $\{1\}$ and $\{2\}$ with equal proportions, and hence, 
\begin{equation}\label{eq:sim:train}
f_{tr}(x) = 0.5 f_1(x) + 0.5 f_2(x). 
\end{equation}
Testing data are generated with the same labels $\{1\}, \{2\}$ but with different percentages,
\begin{equation}\label{eq:sim:test}
f_{te}(x) =  q_1 f_1(x) + (1-q_1) f_2(x),
\end{equation}
where we vary the value of $q_1$ from $(1/2,1/3,\dots,1/10)$ in a decreasing order to present different magnitudes of distribution shift. For each scenario, we generate  training and test data with sample size $500$, and set $X\in \mathbb{R}^{10}$. Testing accuracy is calculated via applying the trained classifier on the testing data. We repeat the simulation $20$ times and report the mean and standard deviation of the testing accuracy. 

{\bf{Setting I: Sparse class boundaries.}} 
We consider both $f_1$ and $f_2$ as mixture density functions with 
\begin{equation}
X\sim 
    \begin{cases}
       \frac{1}{2}N_p(\mu_0,\Sigma) + \frac{1}{2}N_p(-\mu_0, \Sigma) & \; \textrm{if}\; Y=1,  \label{eq:sim:case1:f1} \\
      \frac{1}{2}N_{p}(\mu_1, \Sigma) + \frac{1}{2}N_p(-\mu_1, \Sigma)& \; \textrm{if}\; Y=2, 
    \end{cases}\\
\end{equation}
where $p=10$, $\Sigma=I_{10\times 10}$, $\mu_0=(2,-2,0,\dots,0)^\top\in \mathbb{R}^{10}$, and $\mu_1=(2,2,0,\dots,0)^\top\in \mathbb{R}^{10}$.

\begin{figure}[h!]
\centering
\includegraphics[width=1.08\textwidth]{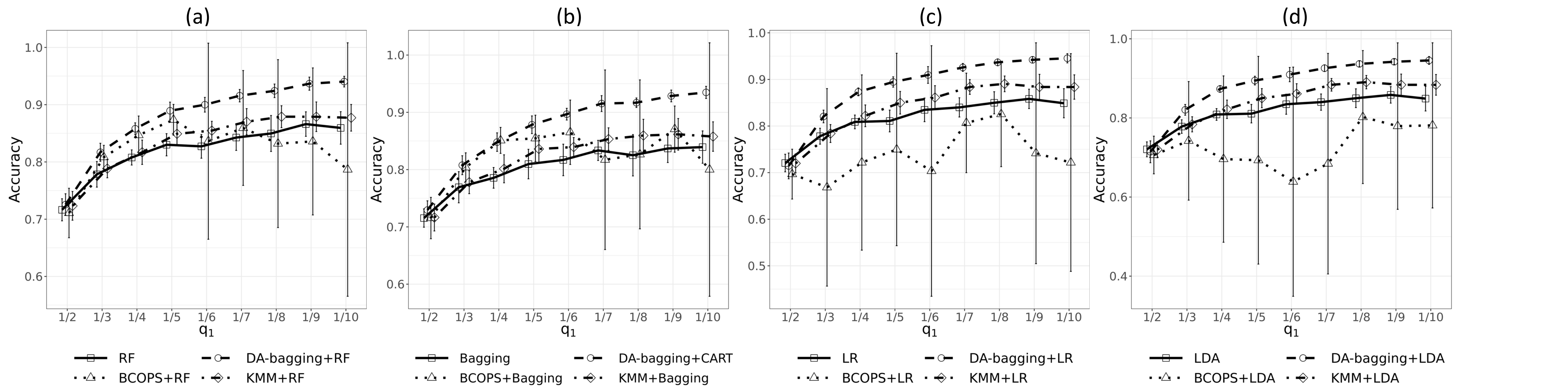}
\caption{Simulation setting I (a)-(d): testing accuracy under different approaches. RF, random forest; Bagging, bootstrap aggregation; LR, logistic regression; LDA, linear discriminant analysis; DA-bagging, domain adaptive bagging; BCOPS, \cite{guan2019prediction}'s method; KMM, \cite{zhang2013domain}'s method; $A+B$ refers to $A$ method equipped with $B$ base classifier.} \label{fig:sim1}
\end{figure}

Figure \ref{fig:sim1} illustrates the test accuracy based on different approaches. \cite{zhang2013domain}'s methods have stable performance for different classifiers but the improvement is not significant over the base classifiers without considering domain adaptation. The performance of \cite{guan2019prediction} varies under different classifiers since the algorithm highly depends on the classifier's ability to detect domain changes. Their method equipped with random forests and bagging significantly improve the test accuracy when $q_1 = 1/2,1/3,1/4,1/5$; however the accuracy drops dramatically when the testing data are highly unbalanced with $q_1<1/5$. For \cite{guan2019prediction} with logistic regression, the accuracy even drops below the baseline classifier. \cite{guan2019prediction} uses a data-splitting strategy to estimate the testing data proportion, which partly explains its unstable performance. Our method has the highest accuracy for different classifiers and performs stably even for very unbalanced data.

{\bf Setting II: Rotated sparse normal.}
\begin{figure}[h!]
\centering
\includegraphics[width=1.08\textwidth]{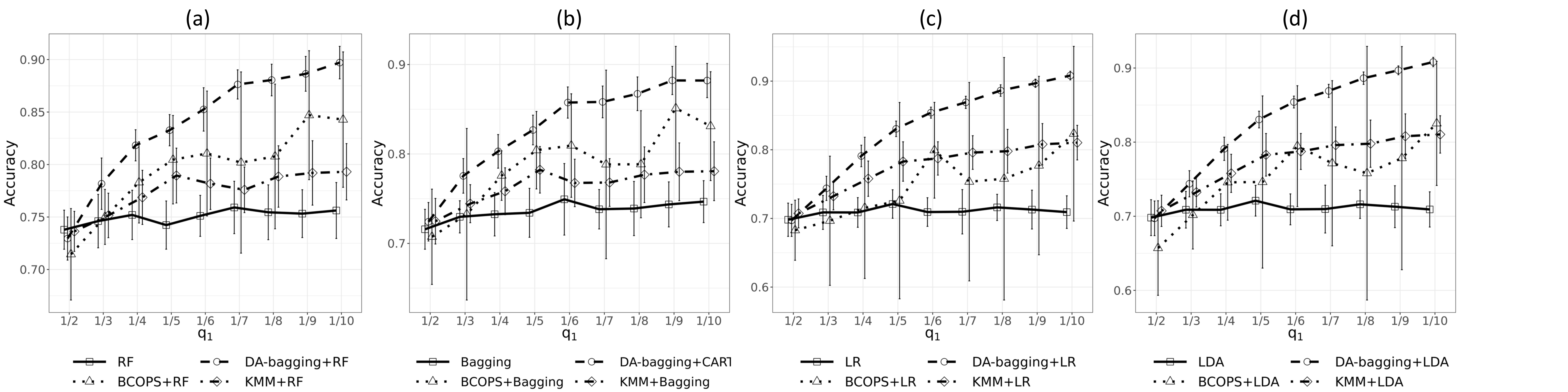}
\caption{Simulation setting II (a)-(d): testing accuracy under different approaches. 
} \label{fig:sim2}
\end{figure}
We consider $f_y(x)$, for $y=1,2$, as multivariate Gaussians with the covariance matrix multiplied by a random rotation matrix, that is, 
\begin{equation}
X\sim 
    \begin{cases}
       N_p(\Omega_p \mu_0, \Omega_p\Sigma_0 \Omega_p^\top) & \; \textrm{if}\; Y=1,  \label{eq:sim:case2:f1} \\
       N_p(\Omega_p \mu_1, \Omega_p\Sigma_1 \Omega_p^\top) & \; \textrm{if}\; Y=2, 
    \end{cases}
\end{equation}
where $p=10$, $\mu_0=(1,1,1,0,\dots,0)^\top$, $\mu_1=(0,0,0,0,\dots,0)^\top$, and $\Omega_p$ is a $p\times p$ rotation matrix sample according to a Haar measure; 
$\Omega_p$ is sampled once and kept as fixed in each replication; $\Sigma_0$ is a block diagonal matrix with two blocks $\Sigma_0^{(1)} = \diag\{\frac{3}{2}\} + \frac{1}{2} \mathbf{1}_3 \mathbf{1}_3^\top$ and $\Sigma_0^{(2)} = \diag\{\frac{1}{2}\} + \frac{1}{2} \mathbf{1}_{p-3} \mathbf{1}_{p-3}^\top$; $\Sigma_1$ is also a block diagonal matrix with two blocks $\Sigma_1^{(1)} =  \diag\{\frac{1}{2}\} + \frac{1}{2} \mathbf{1}_3 \mathbf{1}_3^\top$ and $\Sigma_1^{(2)} = \Sigma_0^{(1)}$.

As shown in Figure \ref{fig:sim2}, all the three domain adaptive methods significantly improve the test accuracy when the testing data differ from the training data. 
Our proposed methods show the highest accuracy and lowest variance when $q_1$ ranges from $1/2$ to $1/10$ for different classifiers.

\subsection{Simulation with anomaly detection}
We accommodate anomalies in the testing data, and consider the model 
\begin{equation}\label{eq:sim:outlier:model}
 f_{tr}(x) = 0.5 f_1(x) + 0.5 f_2(x), \quad  f_{te}(x) = q_1 f_1(x) + q_2 f_2(x) +  \epsilon f_{out}(x).                                           
 \end{equation} 
 We set $f_1(x)$, $f_2(x)$ as in Section \ref{sec:sim:inlier} for two different scenarios. Training data are generated as in Section \ref{sec:sim:inlier} with sample size $n=500$. 
We also generate the testing dataset with sample size $500$ for $q_1$ ranging from $0.9\times(\frac{1}{2},\frac{1}{3}\dots,\frac{1}{10})$, $q_2=0.9-q_1$, and $\epsilon=0.1$,  randomly generating $10\%$ outliers from an alternative distribution. The outlier distribution $f_{out}(x)$ will be specified later. 

 Since \cite{zhang2013domain} is not able to detect the outliers, we compare our proposed method with \cite{guan2019prediction} with four base classifiers considered in Section \ref{sec:sim:inlier}. 
 We set the nominal level as $\alpha=0.1$ for anomaly detection in Algorithm \ref{alg:anomaly}, and compare the empirical type I error and power performance. The empirical type I error is calculated as the percentage of non-outlying data points that are falsely detected, and the empirical power is calculated as the percentage of outliers detected in the testing data. We repeat the simulation $100$ times and report the averaged empirical type I error and power.

{\bf Setting I: Sparse class boundaries with anomalies.} 
In model (\ref{eq:sim:outlier:model}), we generate $X$ from the conditional densities $f_1(x)$ and $f_2(x)$ as in (\ref{eq:sim:case1:f1}). We further generate the anomaly $X_{out}$ with $f_{out}$ the density function of $N_{p}(\mu, \Sigma)$, 
where $p=10$, $\mu = (4,4,0,\dots,0)$ and $\Sigma$ is a diagonal matrix with first two entries as $0.5$ and the remaining $1$. 
Table \ref{tab:1b} compares the type I error and power under different approaches. Domain adaptive bagging controls the type I error at the nominal level while maintaining high detection power; performance is stable under different values of $q_1$. In contrast, \cite{guan2019prediction}'s method has inflated type I error, and the empirical power decreases with $q_1$. The low power can be explained as a sacrifice of their data-splitting procedure.

\begin{table}[h!]
\begin{tabular}{lllrrrrrrrrr}
  \\
& &$q_1/0.9$ & 1/2 & 1/3 & 1/4 & 1/5 & 1/6 & 1/7 & 1/8 & 1/9 & 1/10 \\ 
\multirow{2}{*}{Setting I} &\multirow{2}{*}{DA-bagging}&  type I& 0.06 & 0.06 & 0.05 & 0.05 & 0.05 & 0.05 & 0.05 & 0.04 & 0.05 \\ 
& &  power & 1.00 & 0.99 & 1.00 & 0.99 & 1.00 & 1.00 & 0.99 & 0.99 & 1.00 \\ 
& \multirow{2}{*}{BCOPS+RF} &type I & 0.34 & 0.25 & 0.23 & 0.21 & 0.20 & 0.18 & 0.16 & 0.16 & 0.17 \\
& &  power & 0.97 & 0.96 & 0.96 & 0.93 & 0.91 & 0.92 & 0.86 & 0.87 & 0.86 \\ 
& \multirow{2}{*}{BCOPS+Bagging} & type I & 0.33 & 0.27 & 0.21 & 0.18 & 0.17 & 0.17 & 0.12 & 0.11 & 0.12 \\ 
& & power & 1.00 & 0.99 & 0.96 & 0.95 & 0.95 & 0.93 & 0.87 & 0.86 & 0.81 \\ 
& \multirow{2}{*}{BCOPS+LR}&type I & 0.33 & 0.24 & 0.22 & 0.20 & 0.20 & 0.17 & 0.16 & 0.18 & 0.16 \\ 
& &  power & 0.99 & 0.98 & 0.99 & 0.97 & 0.95 & 0.94 & 0.92 & 0.92 & 0.89 \\ 
& \multirow{2}{*}{BCOPS+LDA}&type I & 0.34 & 0.28 & 0.22 & 0.18 & 0.17 &0.14 & 0.15 & 0.13 & 0.15\\ 
& &  power & 0.99 & 0.98& 0.97& 0.93& 0.94& 0.89& 0.92& 0.87& 0.88 \\ 
\multirow{2}{*}{Setting II}&\multirow{2}{*}{DA-bagging}& type I& 0.06 & 0.05 & 0.04 & 0.04 & 0.04 & 0.04 & 0.04 & 0.04 & 0.04 \\ 
  & &power & 0.58 & 0.55 & 0.52 & 0.53 & 0.57 & 0.56 & 0.53 & 0.55 & 0.54 \\ 
& \multirow{2}{*}{BCOPS+RF}&type I & 0.23 & 0.21 & 0.22 & 0.23 & 0.26 & 0.23 & 0.21 & 0.21 & 0.23 \\ 
  & &power & 0.37 & 0.36 & 0.41 & 0.41 & 0.43 & 0.35 & 0.30 & 0.39 & 0.36 \\ 
& \multirow{2}{*}{BCOPS+Bagging}& type I & 0.19 & 0.19 & 0.18 & 0.16 & 0.17 & 0.18 & 0.20 & 0.18 & 0.18 \\
  & &power & 0.45 & 0.46 & 0.36 & 0.33 & 0.29 & 0.35 & 0.31 & 0.30 & 0.28 \\ 
& \multirow{2}{*}{BCOPS+LR}&type I & 0.21 & 0.19 & 0.21 & 0.22 & 0.23 & 0.22 & 0.20 & 0.21 & 0.21 \\ 
  & &power & 0.31 & 0.29 & 0.33 & 0.35 & 0.37 & 0.31 & 0.24 & 0.32 & 0.31 \\ 
& \multirow{2}{*}{BCOPS+LDA}&type I & 0.18 & 0.16& 0.15 & 0.16 & 0.16 & 0.21&  0.20& 0.18 & 0.20 \\ 
  & &power & 0.31 & 0.32 &  0.30 & 0.33 & 0.25 &  0.34 &  0.35 & 0.34 &  0.32 \\ 
\end{tabular}
\caption{Setting I, II with anomaly: type I error and power based on different approaches. DA-bagging, domain adaptive bagging; BCOPS+RF, BCOPS+Bagging, BCOPS+LR, BCOPS+LDA, refers to \cite{guan2019prediction}'s method with base classifier as random forest, bagging, logistic regression and linear discriminant analysis, respectively. }
\label{tab:1b}
\end{table}

\begin{figure}[h!]
\centering 
\includegraphics[width=1.08\textwidth]{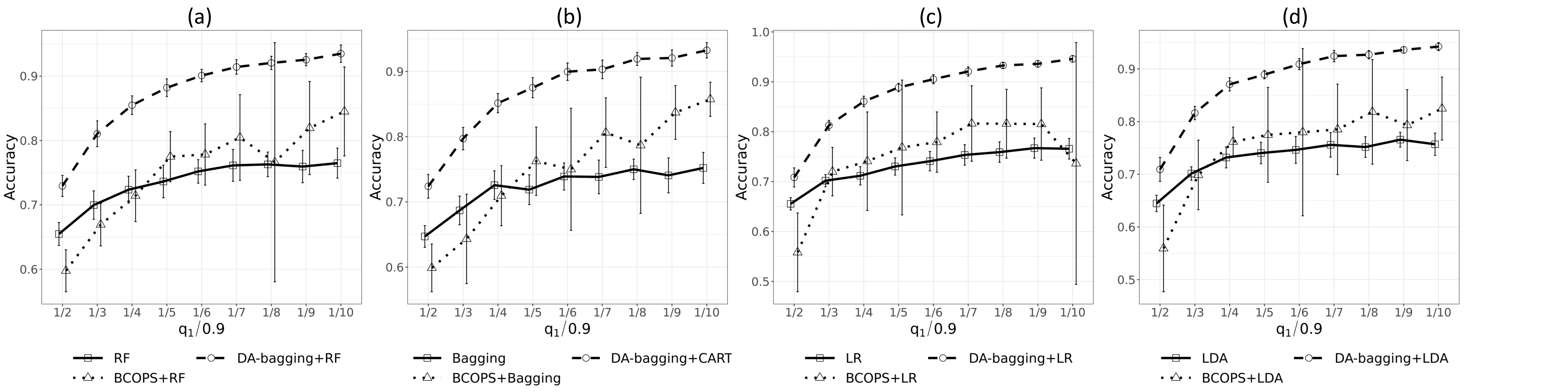}
 \caption{Simulation Setting I (with anomaly) (a)-(d): testing accuracy after anomaly detection under different approaches. RF, random forest; Bagging, bootstrap aggregation; LR, logistic regression; LDA, linear discriminant analysis; DA-bagging, domain adaptive bagging; BCOPS, \cite{guan2019prediction}'s method; KMM, \cite{zhang2013domain}'s method; $A+B$ refers to $A$ method equipped with $B$ base classifier.
 }\label{fig:sim1b}
\end{figure}

{\bf Setting II: Rotated sparse normal with anomalies.} 
Following model (\ref{eq:sim:outlier:model}), we generate $X$ from $f_1(x)$ and $f_2(x)$ as in (\ref{eq:sim:case2:f1}). We further generate anomalies $X_{out}$ from $f_{out}$, the density function of $N_p(\Omega_p \mu_2, \Omega_p I_{p\times p} \Omega_p^\top)$, 
where $\mu_2 = (0,0,2,2,0,\dots,0)$ and $\Omega_p$  is the rotation matrix defined in Setting 2(a).
As shown in Table \ref{tab:1b}, type I error is controlled under the nominal level for domain adaptive bagging. The power is lower compared with Setting I since the outliers' distribution $f_{out}$ is less distinguishable from $f_1$ and $f_2$. The empirical power performance of domain adaptive bagging is still stable for different $q_\ell$. \cite{guan2019prediction}'s approach has inflated type I error, implying over sensitivity in selecting outliers; the power performance of \cite{guan2019prediction}'s method is unsatisfactory and  unstable with respect to different classifiers and class proportions.  

\begin{figure}[h!]
\centering 
\includegraphics[width=1.08\textwidth]{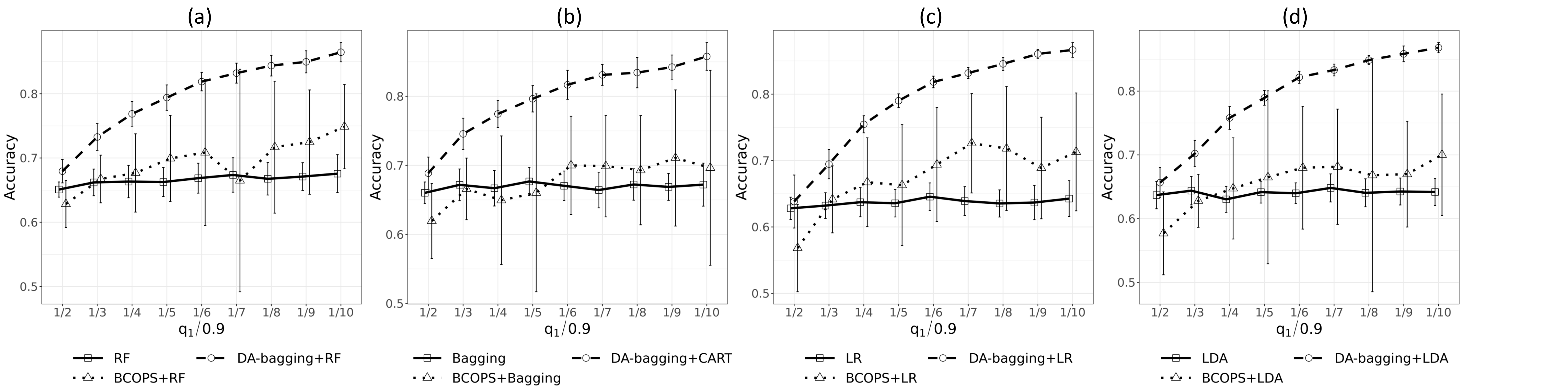}
 \caption{Simulation Setting II (with anomaly) (a)-(d): testing accuracy under different approaches after anomaly detection. RF, random forest; Bagging, bootstrap aggregation; LR, logistic regression; LDA, linear discriminant analysis; DA-bagging, domain adaptive bagging; BCOPS, \cite{guan2019prediction}'s method; KMM, \cite{zhang2013domain}'s method; $A+B$ refers to method  $A$ equipped with base classifier $B$.}\label{fig:sim2b}
\end{figure}

After removing $10\%$ of the data points as possible outliers, we compare different approaches in terms of accuracy on the test sample.  As shown in Figures \ref{fig:sim1b} and \ref{fig:sim2b}, our method still has the highest accuracy for each of different base classifiers. For \cite{guan2019prediction}'s method, the variance increases as $q_1$ decreases. When the  testing data have a similar distribution as the training data, the accuracy of \cite{guan2019prediction} drops below the baseline due to the extra error brought by the inaccurate estimate of the distributional change.

\section{Breast Cancer Data Analysis}\label{sec:real_data}
We apply our method to the Wisconsin
Breast Cancer Dataset available at the University of California,
Irvine machine learning repository. The dataset contains 
nine features of tumors in 699 patients, with 241 malignant and 
458 benign.  Since the data are collected in different time periods, we use the data originally collected in January 1989 with sample size 367 as the training data,  including $200$ benign samples and $167$ malignant samples. 
We use the subsequent 332 data collected from October 1989 to November 1991 as our testing data, which include $258$ benign samples and $74$ malignant samples. The percentage of malignant samples in the testing data is $22.3\%$ which is significantly lower than the percentage, $45.5\%$, of malignant samples in the training data. We use three baseline classifiers including random forest, bagging with classification and regression trees, and logistic regression, as the original classifiers without considering distribution shift. We further equip our approach, \cite{guan2019prediction} and \cite{zhang2013domain}'s methods with these classifiers. 
The tuning parameters are as in the simulation examples. 

To examine the performance of domain adaptive bagging under different training sample sizes, we vary the training data via randomly sampling from the candidate training dataset with sample size $n^{tr}= 50, 100$ and $200$ respectively. The sample size of the test dataset is $n^{te}=332$. Since $n^{te}> n^{tr}$, we first randomly sample $n^{tr}$ data points from the training set, then follow Algorithm S1 for prediction on the test data. We report the test accuracy as the number of wrongly predicted data points in the whole test dataset.  As shown in Table \ref{tab:real2}, our method performs the best compared with other approaches, and the prediction error is stable under different training sample sizes.

\begin{table}
\begin{center}
\begin{tabular}{cccccc}
   & Classifier & DA-bagging & KMM & BCOPS & Original\\
\multirow{4}{*}{$n^{tr} = 50$} & 
 RF        & 5.00 & 9.00 &  15.00 & 13.00 \\ 
 & Bagging & 5.00 & 10.00 & 13.00 & 14.00 \\ 
 & LDA     & 5.00 & 9.00 &  21.00 & 14.00 \\ 
 & LR      & 5.00 & 15.00 & 22.00 & 14.00 \\ 
   &  & DA-bagging & KMM & BCOPS & Original\\
\multirow{4}{*}{$n^{tr} = 100$} &
   RF      & 4.00 & 7.00 & 10.00 & 8.00 \\ 
  & Bagging& 4.00 & 8.00 & 9.00 & 11.00 \\ 
  & LDA    & 4.00 & 9.00 & 15.00 & 7.00 \\ 
  & LR     & 4.00 & 15.00 & 14.00 & 7.00 \\
   &  & DA-bagging & KMM & BCOPS & Original\\
\multirow{4}{*}{$n^{tr} = 200$} & 
  RF       & 4.00 & 5.00 & 5.00 & 8.00 \\ 
  &Bagging & 4.00 & 8.00 & 6.00 & 8.00 \\ 
  &LDA     & 4.00 & 8.00 & 8.00 & 6.00 \\ 
  &LR      & 4.00 & 9.00 & 7.00 & 6.00 \\ 
\end{tabular}\label{tab:real2}
\caption{Breast cancer dataset: number of wrongly predicted data points in the testing set  under different training sample sizes. RF, random forest; Bagging, \cite{breiman1996bagging}'s method; LDA, linear discriminant analysis; LR, logistic regression. DA-bagging, domain adaptive bagging; KMM, kernel mean matching proposed in \cite{zhang2013domain}. BCOPS, \cite{guan2019prediction}'s method. }
\end{center}
\end{table}

\subsection{MNIST Data}
\begin{figure}[H]
\centering 
\includegraphics[width=1.08\textwidth]{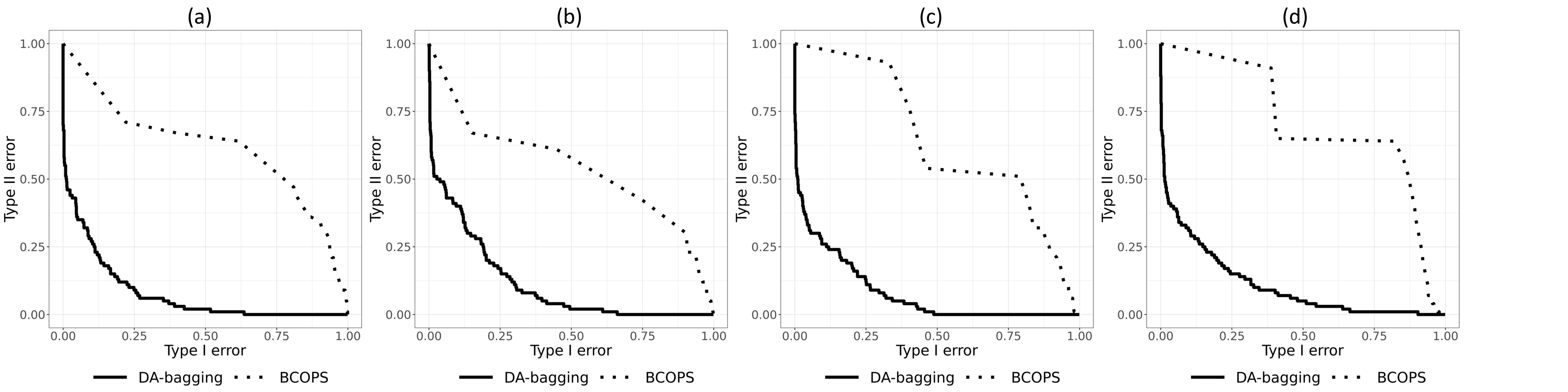}
 \caption{Type I and Type II error for MNIST data set.}\label{fig:real2}
\end{figure}

We analyze the MNIST handwritten digit dataset \citep{lecun2010}. We randomly select $500$ images labeled as $\{5\}$ and $500$ images labeled as $\{6\}$ together as the training data. For the testing data, we randomly sample $900q_1$ images labeled as $\{5\}$ and $900(1-q_1)$ images labeled as $\{6\}$ without overlapping with the training set. We set $q_1$ as $\{1/2, 1/5, 1/10\}$ to present different levels of heterogeneity between the training and  test data.  Then we randomly sample $100$ images with labels in $\{0,1,2\}$ as the new digits unobserved in the training data. We set the nominal level $\alpha=0.1$ in Algorithm 2. Figure \ref{fig:real2} evaluates the empirical type I and type II error. 
Clearly, compared with \cite{guan2019prediction}, domain adaptive bagging has higher power given any fixed type I error.

In addition, we remove $10\%$ of the testing data points with the largest distance to measure for domain adaptive bagging and with the smallest conformal score for \cite{guan2019prediction}'s method. Then we compare the testing accuracy on the remaining data points. In domain adaptive bagging, we choose the number of iterations $T$ based on the threshold $\varepsilon_T=0.01$. We set $B=500$ for both DA-bagging and bagging. 
 We use Random Forest as the baseline classifier. \cite{guan2019prediction}'s method equipped with random forest shows a lower accuracy due to the high error rate in detecting the outliers. For example, when $q_1= 1/2, 1/5$, many inliers are falsely removed, its classification accuracy drops below the accuracy of the random forest baseline classifier without considering distribution shifting.

\begin{table}
\begin{center}
\begin{tabular}{ccccc}
      $q_1$ & DA-bagging+RF &BCOPS+RF&  RF \\
      1/2  &  0.924 & 0.849 & 0.885 \\
      1/5 & 0.940 & 0.860 &  0.881 \\
      1/10 & 0.938 & 0.908 & 0.879 \\
\end{tabular}
\caption{MNIST dataset: testing accuracy. DA-bagging+RF, domain adaptive bagging with random forest as base classifier; BCOPS+RF, \cite{guan2019prediction}'s method with random forest; RF, random forest.}
\label{tab:real3}
\end{center}
\end{table}

\section{Discussion}\label{sec:discussion}

Domain adaptive bagging is a promising general approach for improving classification performance when there is a distributional shift between the training and test data.  Such shifts are common in practice, and methods that fail to adjust can have poor performance.  In this article, we have focused on a particular type of distributional shift, and there are several natural next steps that are of substantial interest.   The first general direction is to accommodate different types of distributional shifts.  For example, instead of only allowing the label proportions to vary, one can also allow the density of the features within each class to vary. In doing this, it is important to include some commonalities between training and test sets.  One possibility is represent the different feature densities with a common set of kernels, but with the weights varying not just due to variation in the label proportions but due to other unknown factors.  

Another possibility, which is particularly natural for high-dimensional and geometrically structured features, is to suppose that there is some lower-dimensional structure in the data.  For example, the features may tend to be concentrated close to a lower-dimensional manifold.  If this lower-dimensional structure tends to be largely preserved between training and test data, then it is natural to leverage on manifold learning or other dimensionality reduction algorithms in constructing relevant distances to be used in applying an appropriate variant of the iterated nearest neighbor sampler within domain adaptive bagging.  

The ideas behind domain adaptive bagging can be applied to related problems in which one wants to improve reproducibility but does not have a specific test set to focus on.  If data are collected under a complex sample survey design and sampling weights are available, then resampling can be modified to produce bootstrap samples from the training data that are population-representative instead of representative of the test data.  Alternatively, if such sampling weights are unavailable, one can generate bootstrap samples that are designed to be highly heterogeneous across covariates groups. Ideally, this would improve generalizability to a variety of distributional shifts that may occur in future test datasets that are as of yet unobserved.  

\appendix 
\section{Appendix: Main Proofs}\label{sec:appendix}

\subsection{A1. Proof of Lemma 1} 
\begin{proof}
Let $B_r(X^{te})$ be the closed ball of radius $r$ centered at $X^{te}$, i.e., $B_r(X^{te}) = \{Z\in \mathbb{R}^p: \|Z-X^{te}\|\leq r\},$ for some metric $\|\cdot\|$ defined on $\mathbb{R}^p$. We first consider a point $X^{te}\in \mathcal{D}_{te}$ such that for any $r>0$, 
\begin{align}
{\pr}_{tr}\{B_r(X^{te})\} = & \pr_{tr}\{Z\in B_r(X^{te})\}=  \int_{B_r(X^{te})} f_{tr}(z)dz
= \sum_{\ell=1}^L \int_{B_r(X^{te})}  p_\ell f_{\ell}(z) dz \nonumber\\
\geq & \sum_{\ell:q_\ell \neq 0} \int_{B_r(X^{te})} p_\ell f_\ell(z) dz \geq \min_{\ell: q_\ell \neq 0}\big\{\frac{p_\ell}{q_\ell}\big\} \sum_{\ell: q_\ell \neq 0} \int_{B_r(X^{te})} q_\ell f_\ell(z) dz \nonumber\\
= &   \min_{\ell: q_\ell \neq 0}\big\{\frac{p_\ell}{q_\ell}\big\} \int_{B_r(X^{te})} f_{te}(z)dz > 0. \label{eq:ball} 
\end{align}
Then, for any $r>0$, given $X^{te}\sim f_{te}$, we have 
\begin{align*}
\pr_{tr}\big(\min_{i\in\{1,\cdots,n\}} \{\|X_i -X^{te} \|\}\geq r\big) = & \prod_{i=1}^n \pr\big(\|X_i- X^{te}\|\geq r, X_i \in \mathcal{D}_{tr} \big) = \prod_{i=1}^n [1-\pr_{tr}\{B_r(X^{te})\}]\\
=& [1-\pr_{tr}\{B_r(X^{te})\}]^n \to 0. 
\end{align*}   

Denote the points $X^{te}$ that do not satisfy (\ref{eq:ball}) as $\bar{\mathcal{X}}$. 
Consider a point $\tilde{X}\in \bar{\mathcal{X}}$, that is, there exists some $\bar{r}$, such that $\pr_{tr}\{B_{\bar{r}}(\tilde{X})\}=0$. By (\ref{eq:ball}), we have $\int_{B_{\bar{r}}(\tilde{X})}f_{te}(z)dz =0$. 
There exists a rational point $a_{\tilde{X}}$, s.t $a_{\tilde{X}} \in B_{\bar{r}/3}(\tilde{X})$. Consequently, there exists a small sphere $B_{\bar{r}/2}(a_{\tilde{X}})$, s.t $B_{\bar{r}/2}(a_{\tilde{X}}) \subset B_{\bar{r}}(\tilde{X})$, and $\int_{B_{\bar{r}/2}(a_{\tilde{X}})}f_{te}(z)dz =0 $, that is, $\pr_{te}(B_{\bar{r}/2}(a_{\tilde{X}}))=0$. 
 Also, $\tilde{X}\in B_{\bar{r}/2}(a_{\tilde{X}})$. Since $a_{\tilde{X}}$ is countable, there is at most a countable set of such spheres that contain the entire $\bar{\mathcal{X}}$. Therefore, $\bar{\mathcal{X}} \subset \cup_{\tilde{X}\in \bar{\mathcal{X}}} B_{\bar{r}/2}(a_{\tilde{X}})$. Then we have $\pr_{te}(\bar{\mathcal{X}})=0$. 
\end{proof}

\subsection{Proof of Theorem \ref{thm:sample:dist}}
\begin{proof}
To simplify notation, for any $X^{te}\sim f_{te}(x)$, define its nearest neighbor in $D^{b*,(t)}_{tr}$ as $X_{(1)}$, and the corresponding label of $X_{(1)}$ is $Y_{(1)}$. 
 Denote $\pr^{(t)}_{te}(Y_{(1)}=1) $ as the probability that for $X\sim f_{te}$, its nearest neighbor in $D^{b*, (t)}_{tr}$ is labeled as class 1. Here we use the unified symbol  $D^{(t)}_{tr}$ to represent  $D^{b*,(t)}_{tr}$ for $b=1,\dots, B$. Then we show that $\pr^{(t)}_{te}(Y_{(1)}=1) $ approaches to $q_1$ as $t$ increases. 
Define $p_1^{(t)} = \pr^{(t)}_{te}(Y_{(1)} = 1)$. 
Without loss of generality, suppose $p_1 < q_1$. Then we have 
\begin{align}
\pr^{(1)}_{te}(Y_{(1)} = 1) = & \bE \mathds{1}_{\{Y_{(1)}=1\}} = \bE_{x\sim f_{te}} (\bE [\mathds{1}_{\{Y_{(1)}=1\}} | X^{te}=x ] ) \nonumber\\
= & \bE_{x\sim f_{te}} [\pr \{Y_{(1)} = 1 | X^{te} =x \}] = \bE_{x\sim f_{te}} \frac{p_1f_1(X_{(1)})}{p_1 f_1(X_{(1)}) + (1-p_1)f_2(X_{(1)})}  \nonumber\\
= & \bE_{x\sim f_{te}} \frac{p_1f_1(x)}{p_1 f_1(x) + (1-p_1)f_2(x)} %
= p_1 \int \frac{f_1(x)f_{te}(x)}{p_1 f_1(x) + (1-p_1)f_2(x)} dx \nonumber\\
= & p_1 \int \frac{q_1 f_1(x)+(1-q_1)f_2(x)}{p_1 f_1(x) + (1-p_1)f_2(x)} f_1(x) dx %
=  p_1 \bE_{x\sim f_1} \frac{q_1 f_1(x)+(1-q_1)f_2(x)}{p_1 f_1(x) + (1-p_1)f_2(x)} \nonumber\\
= & p_1 \bE_{x\sim f_1} \frac{q_1 + (1-q_1)f_2(x)/f_1(x)}{p_1 + (1-p_1) f_2(x)/f_1(x)} %
\geq  p_1 \frac{q_1 + (1-q_1)\bE_{x\sim f_1}f_2(x)/f_1(x)}{p_1 + (1-p_1) \bE_{x\sim f_1}f_2(x)/f_1(x)} \label{eq:proof:jensen}%
\end{align} 
where the first expectation is with respect to both $X^{te}\sim f_{te}(x)$ and $(X_{(1)}, Y_{(1)})$ in $\mathcal{D}_{tr}$; the fifth equation is due to $X_{(1)} \to X^{te}$ with probability 1 by Lemma 1 %
and the Lipschitz property of $\frac{p_1f_1(x)}{p_1f_1(x)+(1-p_1)f_2(x)}$;  the inequality in (\ref{eq:proof:jensen}) is due to the fact that $\psi(z) = \frac{q_1+ (1-q_1)z }{p_1+(1-p_1)z}$ is convex, and applying the Jensen's inequality. Note that the equality holds in the last step if and only if $\pr_{f_1}\{ X: f_1(X) = f_2(X)\} = 1$. By the assumption that $\pr_{f_1}\{ X: f_1(X) = f_2(X)\} = 0$, we have 
$$
\pr^{(1)}_{te}(Y_{(1)} = 1) > p_1 \frac{q_1 + (1-q_1)\bE_{x\sim f_1}f_2(x)/f_1(x)}{p_1 + (1-p_1) \bE_{x\sim f_1}f_2(x)/f_1(x)} = p_1. 
$$

Denote the sampling data as $\mathcal{D}_{tr}^{(1)}$, and for $(X_i, Y_i)\in \mathcal{D}_{tr}^{(1)}$, denote the conditional density of $X_i$ given $Y_i=\ell$ for $\ell=1,2$ as $f^{(1)}(x|Y_i=\ell)$. By Algorithm 2, we have 
$f^{(1)}(x|Y=1) = f_1(x)$ and $f^{(1)}(x|Y=2)= f_2(x)$. The marginal distribution of $X_i$ in $\mathcal{D}_{tr}^{(1)}$, i.e., $f^{(1)}(x)$ has the expression 
\begin{align*}
f^{(1)}(x) = & \pr^{(1)}_{te}(Y_{(1)} = 1) f^{(1)}(x|Y=1) + \pr^{(1)}_{te}(Y_{(1)} = 2) f^{(1)}(x|Y=2)\\
= & p_1^{(1)}f_1(x)  + (1-p_1^{(1)}) f_2(x),
\end{align*}
with $p_1^{(1)}> p_1$. 

Next, we show that $ p_1^{(t)}$ increases as $t$ increases when $p_1^{(t)}< q_1$, then converges to $q_1$ as $t\to \infty$. 
At the $t$-th iteration, denote the sampled data as $\mathcal{D}_{tr}^{(t)}$. For any $(X_i, Y_i)\in \mathcal{D}_{tr}^{(t)}$, the prior distribution of $Y_i$ is $\pr(Y_i=1) = \pr_{te}^{(t)}(Y_{(1)} = 1):= p_1^{(t)}$, the marginal distribution of $X_i$ can be written as 
$$
f^{(t)}(x) = p_1^{(t)} f_1(x) + (1-p_1^{(t)}) f_2(x).
$$
If $p_1<p_1^{(t)}< q_1$, then using the above strategy but replacing $\mathcal{D}_{tr}$ and $f_{tr}$ by $\mathcal{D}_{tr}^{(t)}$ and $f^{(t)}$, at the $(t+1)-$th iteration, we have $p_1^{(t+1)}> p_1^{(t)}$.

If $p_1^{(t)}> q_1$, at the $(t+1)$-th iteration, for any test data $X^{te}$, its nearest neighbor at $\mathcal{D}_{tr}^{(t)}$ is denoted as $X_{(1)}$, and the corresponding label of $X_{(1)}$ is $Y_{(1)}$. 
\begin{align}
& \pr^{(t+1)}_{te}(Y_{(1)} = 1)\\
= & \bE_{x\sim f_{te}} [\pr\{Y_{(1)} = 1 | X^{te} =x \}] = \bE_{x\sim f_{te}} \frac{p^{(t)}_1f^{(t)}_1(X_{(1)})}{p^{(t)}_1 f_1(X_{(1)}) + (1-p^{(t)}_1)f_2(X_{(1)})}  \nonumber\\
= & \bE_{x\sim f_{te}} \frac{p^{(t)}_1f_1(x)}{p^{(t)}_1 f_1(x) + (1-p^{(t)}_1)f_2(x)} 
= p^{(t)}_1 \int \frac{f^{(t)}_1(x)f_{te}(x)}{p^{(t)}_1 f_1(x) + (1-p^{(t)}_1)f_2(x)} dx \nonumber\\
= & p^{(t)}_1 \int \frac{q_1 f_1(x)+(1-q_1)f_2(x)}{p^{(t)}_1 f_1(x) + (1-p^{(t)}_1)f_2(x)} f_1(x) dx %
=  p^{(t)}_1 \bE_{x\sim f_1} \frac{q_1 f_1(x)+(1-q_1)f_2(x)}{p^{(t)}_1 f_1(x) + (1-p^{(t)}_1)f_2(x)} \nonumber\\
= & p^{(t)}_1 \bE_{x\sim f_1} \frac{q_1 + (1-q_1)f_2(x)/f_1(x)}{p^{(t)}_1 + (1-p^{(t)}_1) f_2(x)/f_1(x)} 
<  p^{(t)}_1 \frac{q_1 + (1-q_1)\bE_{x\sim f_1}f_2(x)/f_1(x)}{p^{(t)}_1 + (1-p^{(t)}_1) \bE_{x\sim f_1}f_2(x)/f_1(x)} 
=  p_1^{(t)}, \label{eq:proof:jensen:concave} 
\end{align} 
where (\ref{eq:proof:jensen:concave}) is due to Jensen's inequality and the function $\psi(z)= \frac{q_1 + (1-q_1)z}{p+ (1-p)z}$ is concave when $p>q_1$. Therefore, whenever $p_1^{(t)}> q_1$, in the next iteration, the proportion $p_1^{(t+1)}$ will decrease. 

 If $\lim_{t\to\infty }p_1^{(t)}= q_1$, then 
\begin{align*}
\lim_{t\to\infty }p_1^{(t+1)}= & \lim_{t\to \infty }\pr^{(t+1)}_{te}(Y_{(1)} = 1) = \bE_{x\sim f_{te}} \frac{p_1^{(t)}f_1(x)}{p_1^{(t)} f_1(x) + (1-p_1^{(t)})f_2(x)} \\
=& \lim_{t\to\infty}p_1^{(t)} \bE_{x\sim f_1}\frac{q_1 f_1(t) + (1-q_1)f_1(t)}{p_1^{(t)} f_1(x) + (1-p_1^{(t)})f_2(x)} 
= \lim_{t\to \infty}p_1^{(t)}=q_1.  
\end{align*}
Therefore, $\lim_{t \to \infty} \pr^{(t+1)}_{te} (Y_{(1)} = 1 ) = q_1.$ 
On the other hand, according to our sampling procedure, in $\mathcal{D}_{tr}^{(T)}$, the conditional density of $X_i|Y_i = \ell$ is $f^{(T)}(x|Y=\ell) = f_\ell$ for $\ell=1,2$. Hence the marginal density of $X_i$ in $\mathcal{D}_{tr}^{(T)}$ is $f^{(T)}(x) = q_1 f_1(x) + (1-q_1)f_2(x)$.
\end{proof}
\subsection{Proof of Theorem 2} %
\begin{proof}
We now adapt the proof of Theorem 2 in \cite{cannings2017random} to our case. Define $\eta(x) = \pr(Y^{te}=1\mid X^{te}=x)$, then the joint density of $(X,Y)$ for testing set is $f(x,y) = f_{te}(x)\eta(x) + f_{te}(x)\{1-\eta(x)\}$. Then we have 
\begin{align*}
& R(C_n^{DA}) - R(C^{Bayes}) \\
= & \int \{\eta(x)[\mathds{1}_{\{C_n^{DA}(x)=2\}} - \mathds{1}_{\{C^{Bayes}(x)=2\}}] + \{1-\eta(x)\}[\mathds{1}_{\{C_n^{DA}(x)=1\}} - \mathds{1}_{\{C^{Bayes}(x)=1\}}]\}f_{te}(x)dx\\
= & \int \big[ |2\eta(x)-1||\mathds{1}_{\{\Lambda_n(x)< 1/2\}} - \mathds{1}_{\{\eta(x)<1/2\}}|\big]f_{te}(x)dx\\
= & \int \big[ |2\eta(x)-1| \mathds{1}_{\{\Lambda_n(x)\geq 1/2\}}\mathds{1}_{\{\eta(x)<1/2\}} + |2\eta(x)-1|\mathds{1}_{\{\Lambda_n(x)< 1/2 \}}\mathds{1}_{\{\eta(x)\geq 1/2\}}\big] f_{te}(x) dx \\
\leq & \int \big[2|2\eta(x)-1|\Lambda_n(x)\mathds{1}_{\{\eta(x)<1/2\}} + 2|2\eta(x)-1|\{1-\Lambda_n(x)\}\mathds{1}_{\{\eta(x)\geq 1/2\}} \big] f_{te}(x) dx.
\end{align*}
Note that $\bE \{R(C_n^{DA}) - R(C^{Bayes})\} = \bE [\bE \{R(C_n^{DA})\mid \mathcal{D}_{tr}, \mathcal{D}_{te}\}] - R(C^{Bayes})$. 
 Conditioning on $(\mathcal{D}_{tr}, \mathcal{D}_{te})$, $\xi_1,\dots, \xi_B$ are independent and identically distributed. Hence, $C_n^{\xi_b}$ for $b=1,\dots, B$ are independent and identically distributed. Therefore, 
\begin{align*}
 & \bE \{R(C_n^{DA})\mid \mathcal{D}_{tr}, \mathcal{D}_{te} \} - R(C^{Bayes})\\
 =&  \bE \big(\int \big[2|2\eta(x)-1|\mathds{1}_{\{C_n^{\xi_1}(x)=1\}}\mathds{1}_{\{\eta(x)< 1/2\}} \\
 &  \quad + 2|2\eta(x)-1|\mathds{1}_{\{C_n^{\xi_1}(x)=2\}}\mathds{1}_{\{\eta(x)\geq 1/2\}} \big] f_{te}(x) dx \mid \mathcal{D}_{tr}, \mathcal{D}_{te}\big)\\
 \leq & 2\bE \big(\int |2\eta(x)-1||\mathds{1}_{\{C_n^{\xi_1}(x)=2\}} - \mathds{1}_{\{\eta(x)< 1/2\}}| f_{te}(x)dx \mid \mathcal{D}_{tr}, \mathcal{D}_{te} \big) \\
 = & 2\big[\bE \{R(C_n^{\xi_1})\}-R(C^{Bayes}) \mid \mathcal{D}_{tr}, \mathcal{D}_{te}\big].       
\end{align*}
Then we have $\bE R(C_n^{DA}) - R(C^{Bayes}) = \bE [\bE \{R(C_n^{DA})\mid \mathcal{D}_{tr}, \mathcal{D}_{te} \}- R(C^{Bayes}) \leq 2 \big[\bE  \{R(C_n^{\xi_1})\}-R(C^{Bayes}) \big]. $
\end{proof}

\subsection{Proof of Theorem 3}
\begin{proof}
Conditional on $(\cD_{tr}, \cD_{te})$, $\xi_1,\dots, \xi_B$ are independent and identically distributed. For any pair $(X,Y)$ with $\pr(Y=\ell)=q_\ell$ and $f(X=x|Y=\ell) = f_\ell(x)$, the test error of the ensemble classifier can be written as 
\begin{align*}
& \bE \{R(C_n^{DA}) \mid \mathcal{D}_{tr}, \mathcal{D}_{te}\} \\
= & \bE [q_1 \int_{\mathbb{R}^p} \mathds{1}_{\{C_n^{DA}(x)=2\}} f_1(x) dx + q_2 \int_{\mathbb{R}^p}  \mathds{1}_{\{C_n^{DA}(x)=1\}}f_2(x) dx]\\
= & \bE[ q_1 \int_{\mathbb{R}^p}  \mathds{1}_{\{\Lambda_n(x)<1/2\}}f_1(x) dx + q_2 \int_{\mathbb{R}^p}  \mathds{1}_{\{\Lambda_n(x)\geq 1/2\}} f_2(x)dx]\\
= & q_1 \int_{\mathbb{R}^p}  \pr\{\Lambda_n(x) < 1/2 \mid \mathcal{D}_{tr}, \mathcal{D}_{te}\} f_1(x)dx + q_2 \int_{\mathbb{R}^p}  \pr\{\Lambda_n(x) \geq 1/2 \mid \mathcal{D}_{tr}, \mathcal{D}_{te}\}f_2(x) dx, 
\end{align*}
where the final equality follows by Fubini's theorem. Let $U_{b} := \mathds{1}_{\{C_n^{\xi_b}(X)=1\}}$ for $b=1,\dots, B$. Then, conditional on $\mu_n(X)=\theta\in[0,1]$, the random variables $U_1, \dots, U_B$ are independent, each having a Bernoulli($\theta$) distribution. Denote $\mathcal{L}_{\mu_n,1}$ and $\mathcal{L}_{\mu_n,2}$ as the distribution function of $\mu_n(X)\mid Y=1$ and $\mu_n(X) \mid Y=2$. That is, $\mathcal{L}_{\mu_n,\ell}$ is short for $\mathcal{L}_{\mu_n,\ell}( t \mid  \mathcal{D}_{tr}, \mathcal{D}_{te}, Y=\ell)$ for $\ell=1,2$. Then we have 
\begin{align*}
\int_{\mathbb{R}^p}  \pr\{\Lambda_n(x) < 1/2 \mid \mathcal{D}_{tr}, \mathcal{D}_{te}\} f_1(x)dx = & \int_{[0,1]} \pr\{\frac{1}{B}\sum_{b=1}^B U_b < 1/2 \mid \mu(X)= \theta\} d\mathcal{L}_{\mu_n,1}(\theta) \\=& \int_{[0,1]} \pr(T< B/2) d\mathcal{L}_{\mu_n,1}(\theta), 
\end{align*}
where $T$ denotes a random variable following Binomial distribution with parameters $B, \theta$; that is, $T\sim \textrm{Bin}(B,\theta)$. Similarly, we can write 
$$
 \int_{\mathbb{R}^p}  \pr\{\Lambda_n(x) \geq 1/2\mid \mathcal{D}_{tr}, \mathcal{D}_{te}\}f_2(x) dx = 1 - \int_{[0,1]} \pr(T< B/2) d\mathcal{L}_{\mu_n,2}(\theta).
$$ 
Therefore, 
$$
\bE\{R(C_n^{DA}) \mid \mathcal{D}_{tr}, \mathcal{D}_{te}\} = q_2 + \int_{[0,1]} \pr(T< B/2) d\mathcal{L}_{\mu_n,2}(\theta). 
$$
Then we aim to show that 
\begin{equation}\label{pf:thm:alg}
  \int_{[0,1]} \{\pr(T < B/2) - \mathds{1}_{\{\theta < 1/2\}}\} d\{q_1 \mathcal{L}_{n,1}(\theta) - q_2 \mathcal{L}_{n,2}(\theta)\}
= \gamma_n,
\end{equation}
 where $$\gamma_n = (1/2 - (B/2 - \floor{B/2}))\{q_1 g_{n,1}(1/2) - q_2 g_{n,2}(1/2)\}+ \frac{1}{8} \{q_1 \dot{g}_{n,1}(1/2)-q_2 \dot{g}_{n,2}(1/2)\},$$ with $g_{n,\ell}$ and $ 
\dot{g}_{n,\ell}$ the first and second order derivative of $\mathcal{L}_{\mu_n,\ell}( t \mid \mathcal{D}_{tr}, \mathcal{D}_{te}, Y^{te}=\ell)$ for $\ell=1,2$. The proof of (\ref{pf:thm:alg}) requires a one-term Edgeworth expansion to the binomial distribution function; refer to the  proof of Theorem 1 in \cite{cannings2017random}.  
\end{proof}

\subsection{Proof of Theorem 4} %
Before the proof of Theorem 4, we first introduce a useful lemma to bound the difference between $d_\ell(x)$ and $\widehat{d}_\ell(x)$. 
\begin{lemma}\label{le:pre:consist}
Suppose Assumptions 2 and 3 hold, with $\delta$ specified in Assumption 2 (b). 
Denote the support of $\pr_\ell$ and $\pr_0$ as $S_\ell$ and $S_0$. 
Then with probability at least $1-\delta$, for any $x\in S_\ell \cup S_0$, 
$$
\sup_x |d_\ell(x) - \widehat{d}_\ell (x)| \leq C \beta_{n_\ell}(\beta_{n_\ell} + \sqrt{m_{\ell}}).
$$
\end{lemma}
Defining the event $\mathcal{E}_\ell = \{\sup_x |d_\ell(x) - \widehat{d}_\ell (x)| \leq C \beta_{n_\ell}(\beta_{n_\ell} + \sqrt{m_\ell})\}$, Lemma \ref{le:pre:consist} guarantees that the event $\mathcal{E}_\ell$ holds with high probability. That is, $\widehat{d}_\ell$ is a consistent estimate of $d_\ell$ for all $x$ in $S_\ell \cup S_0$. The proof of Lemma \ref{le:pre:consist} refers to Theorem 3.5 in \cite{gu2019statistical}. 

Next, we are ready to prove Theorem 4. 
\begin{proof}
We first prove Theorem 4 (a). 
Under $H_0$, for any $X^{te}\in \mathcal{D}_{te}$, denote the corresponding label as $Y^{te}$. Suppose $Y^{te}=\ell$, i.e., $\ell$ is the true label of $X^{te}$. 
\begin{align*}
& \pr\{T(X^{te})>c \mid  Y^{te} =\ell \}\\
 = & \pr[\{\widehat{d}_1(X^{te}) > c_1\}\cap \dots \cap \{\widehat{d}_L(X^{te}) > c_L\} \mid Y^{te}=\ell ]\\
\leq &  \pr[\widehat{d}_{\ell} (X^{te}) > c_{\ell} \mid Y^{te}=\ell]\\ 
= & \pr[d_{\ell}(X^{te}) > c_{\ell} - \{\widehat{d}_{\ell}(X^{te}) - d_{\ell}(X^{te})\}\mid  Y^{te}=\ell] \\
\leq  & \pr[d_{\ell}(X^{te}) > c_{\ell} - \{\widehat{d}_{\ell}(X^{te}) - d_{\ell}(X^{te})\} \mid Y^{te}=\ell, \mathcal{E}_{\ell}]\pr(\mathcal{E}_{\ell}) + \pr(\mathcal{E}_{\ell}^c)\\
\leq &  \pr\{d_{\ell}(X^{te}) > c_\ell - C \beta_{n_\ell} (\beta_{n_\ell} + \sqrt{m_{\ell}}) \mid Y^{te}=\ell \} (1-\delta) + \delta
\leq \alpha + o(1),
\end{align*}
where the last step is based on Lemma \ref{le:pre:consist} and $\beta_{n_\ell} (\beta_{n_\ell} + \sqrt{m_{\ell}})=o(c_\ell)$ as $n_\ell\to \infty$. 

We next prove the power in Theorem 4 (b). Note that
\begin{align*}
& \pr[T(X^{te})>c \mid  Y^{te}\not\in \{1,\dots, L\}] \\
= & \pr[\{\widehat{d}_1(X^{te}) > c_1\}\cap \dots \cap \{\widehat{d}_L(X^{te}) > c_L\} \mid Y^{te}\not\in \{1,\dots, L\} ]\\
\geq & 1 - \sum_{\ell=1}^L \pr_\ell [\widehat{d}_\ell (X^{te}) \leq c_\ell \mid Y^{te}\not\in \{1,\dots, L\}]. 
\end{align*}
We provide an upper bound for $\pr [\widehat{d}_\ell (X^{te}) \leq c_\ell \mid Y^{te}\not\in \{1,\dots, L\}] $. It is sufficient to prove $\inf_{X^{te}\sim f_{out}} d_\ell (X^{te}) > c_\ell + C \beta_{n_\ell} (\beta_{n_\ell} + \sqrt{m_\ell})$ under event $\mathcal{E}_\ell$. Note that 
\begin{align*}
\inf_{X^{te}\sim f_{out}} d_\ell (X^{te}) = &  \inf_{X^{te}\sim f_{out}} \frac{1}{m_\ell}\int_0^{m_\ell} r^2_{\pr_\ell, t}(X^{te}) dt\\
 \geq & \inf_{X^{te}\sim f_{out}} \frac{1}{m_\ell} \int_{\epsilon}^{m_\ell} r^2_{\pr_\ell, t}(X^{te}) dt \geq \frac{m_\ell-\epsilon}{m_\ell} M > c_\ell +  C \beta_{n_\ell} (\beta_{n_\ell} + \sqrt{m_\ell}),
\end{align*}
where the last step is due to Assumption 2 (a). 
Therefore, 
\begin{align*}
& \pr[\widehat{d}_\ell(X^{te}) <  c_\ell \mid Y^{te}\not\in \{1,\dots, L\}]\\
 \leq & \pr[\widehat{d}_\ell (X^{te}) < c_\ell \mid \mathcal{E}_\ell, \; Y^{te}\not\in \{1,\dots, L\}]\pr(\mathcal{E}_\ell) + \pr(\mathcal{E}_\ell^c) \\
\leq & \pr[ \widehat{d}_\ell (X^{te}) < c_\ell \mid \mathcal{E}_\ell, \; Y^{te}\not\in \{1,\dots, L\}]\pr(\mathcal{E}_\ell) + \pr(\mathcal{E}_\ell^c)\\
\leq & \pr[d_\ell (X^{te}) - C \beta_{n_\ell} (\beta_{n_\ell} + \sqrt{m_{\ell}})< c_\ell \mid \mathcal{E}_\ell, \; Y^{te}\not\in \{1,\dots, L\}]\pr(\mathcal{E}_\ell) + \pr(\mathcal{E}_\ell^c)\\
\leq & \pr(\mathcal{E}^c) \leq \delta.
\end{align*}
Then we have 
$$
\pr[T(X^{te})>c \mid Y^{te}\not\in \{1,\dots, L\}] \geq  1 - \sum_{\ell=1}^L \pr_\ell [\widehat{d}_\ell (X^{te}) \leq c_\ell \mid Y^{te}\not\in \{1,\dots, L\}] \geq 1-L\delta.  
$$
\end{proof}

\bibliographystyle{abbrvnat}
\bibliography{reference}

\begin{thebibliography}{17}
\providecommand{\natexlab}[1]{#1}
\providecommand{\url}[1]{\texttt{#1}}
\expandafter\ifx\csname urlstyle\endcsname\relax
  \providecommand{\doi}[1]{doi: #1}\else
  \providecommand{\doi}{doi: \begingroup \urlstyle{rm}\Url}\fi

\bibitem[Breiman(1996)]{breiman1996bagging}
L.~Breiman.
\newblock Bagging predictors.
\newblock \emph{Machine Learning}, 24\penalty0 (2):\penalty0 123--140, 1996.

\bibitem[Breiman(2001)]{breiman2001random}
L.~Breiman.
\newblock Random forests.
\newblock \emph{Machine Learning}, 45\penalty0 (1):\penalty0 5--32, 2001.

\bibitem[Cannings and Samworth(2017)]{cannings2017random}
T.~I. Cannings and R.~J. Samworth.
\newblock Random-projection ensemble classification.
\newblock \emph{Journal of the Royal Statistical Society: Series B (Statistical
  Methodology)}, 79\penalty0 (4):\penalty0 959--1035, 2017.

\bibitem[Chan and Ng(2005)]{chan2005word}
Y.~S. Chan and H.~T. Ng.
\newblock Word sense disambiguation with distribution estimation.
\newblock \emph{Proceedings of the Nineteenth International Joint Conference on
  Artificial Intelligence}, 5:\penalty0 1010--1015, 2005.

\bibitem[Chazal et~al.(2011)Chazal, Cohen-Steiner, and
  M{\'e}rigot]{chazal2011geometric}
F.~Chazal, D.~Cohen-Steiner, and Q.~M{\'e}rigot.
\newblock Geometric inference for probability measures.
\newblock \emph{Foundations of Computational Mathematics}, 11\penalty0
  (6):\penalty0 733--751, 2011.

\bibitem[Chazal et~al.(2017)Chazal, Fasy, Lecci, Michel, Rinaldo, Rinaldo, and
  Wasserman]{chazal2017robust}
F.~Chazal, B.~Fasy, F.~Lecci, B.~Michel, A.~Rinaldo, A.~Rinaldo, and
  L.~Wasserman.
\newblock Robust topological inference: Distance to a measure and kernel
  distance.
\newblock \emph{The Journal of Machine Learning Research}, 18\penalty0
  (1):\penalty0 5845--5884, 2017.

\bibitem[Cochran and Rubin(1973)]{cochran1973controlling}
W.~G. Cochran and D.~B. Rubin.
\newblock Controlling bias in observational studies: A review.
\newblock \emph{Sankhy{\=a}: The Indian Journal of Statistics: Series A},
  35:\penalty0 417--446, 1973.

\bibitem[Gu et~al.(2019)Gu, Akoglu, and Rinaldo]{gu2019statistical}
X.~Gu, L.~Akoglu, and A.~Rinaldo.
\newblock Statistical analysis of nearest neighbor methods for anomaly
  detection.
\newblock \emph{Advances in Neural Information Processing Systems},
  32:\penalty0 10923--10933, 2019.

\bibitem[Guan and Tibshirani(2019)]{guan2019prediction}
L.~Guan and R.~Tibshirani.
\newblock Prediction and outlier detection: a distribution-free prediction set
  with a balanced objective.
\newblock \emph{arXiv preprint arXiv:1905.04396}, 2019.

\bibitem[Hall et~al.(2008)Hall, Park, and Samworth]{hall2008choice}
P.~Hall, B.~U. Park, and R.~J. Samworth.
\newblock Choice of neighbor order in nearest-neighbor classification.
\newblock \emph{The Annals of Statistics}, 36\penalty0 (5):\penalty0
  2135--2152, 2008.

\bibitem[Heckman(1990)]{heckman1990varieties}
J.~Heckman.
\newblock Varieties of selection bias.
\newblock \emph{The American Economic Review}, 80\penalty0 (2):\penalty0
  313--318, 1990.

\bibitem[LeCun et~al.(2010)LeCun, Cortes, and Burges]{lecun2010}
Y.~LeCun, C.~Cortes, and C.~Burges.
\newblock Mnist handwritten digit database.
\newblock \emph{\url{http://yann.lecun.com/exdb/mnist/}}, 2010.

\bibitem[Lipton et~al.(2018)Lipton, Wang, and Smola]{lipton2018detecting}
Z.~C. Lipton, Y.~Wang, and A.~J. Smola.
\newblock Detecting and correcting for label shift with black box predictors.
\newblock \emph{Proceedings of the 35th International Conference on Machine
  Learning}, 80:\penalty0 3128--3136, 2018.

\bibitem[Lopes(2020)]{lopes2020estimating}
M.~E. Lopes.
\newblock Estimating a sharp convergence bound for randomized ensembles.
\newblock \emph{Journal of Statistical Planning and Inference}, 204:\penalty0
  35--44, 2020.

\bibitem[Storkey(2009)]{storkey2009training}
A.~Storkey.
\newblock When training and test sets are different: Characterizing learning
  transfer.
\newblock \emph{Dataset Shift in Machine Learning}, pages 3--28, 2009.

\bibitem[Tucker(2010)]{tucker2010selection}
J.~W. Tucker.
\newblock Selection bias and econometric remedies in accounting and finance
  research.
\newblock \emph{Journal of Accounting Literature}, 29:\penalty0 31--57, 2010.

\bibitem[Zhang et~al.(2013)Zhang, Sch{\"o}lkopf, Muandet, and
  Wang]{zhang2013domain}
K.~Zhang, B.~Sch{\"o}lkopf, K.~Muandet, and Z.~Wang.
\newblock Domain adaptation under target and conditional shift.
\newblock \emph{International Conference on Machine Learning}, 28:\penalty0
  819--827, 2013.

\end{thebibliography}
\end{document}